\documentclass[aip,amsmath,amssymb,reprint,nofootinbib]{revtex4-1}
\usepackage{amsmath}
\usepackage{amsfonts}
\usepackage{graphicx}
\usepackage{epstopdf}
\usepackage{bbold}
\usepackage{array}
\usepackage{empheq}
\usepackage{subcaption}
\usepackage[bottom]{footmisc}
\usepackage{textcomp}
\usepackage[shortlabels]{enumitem}
\usepackage{overpic}
\usepackage{xcolor}
\definecolor{blue}{rgb}{0, 0, 1}
\definecolor{red}{rgb}{1, 0, 0}
\definecolor{green}{rgb}{0.011, 0.128, 0.064}
\usepackage{ragged2e} % for the \justifying macro
\DeclareCaptionJustification{justified}{\justifying}
\begin{document}

\title{SPLEND1D, a reduced one-dimensional model to investigate the physics of plasma detachment}

\author{O. F\'evrier}
\email[]{olivier.fevrier@epfl.ch}
\affiliation{Swiss Plasma Center (SPC), \'Ecole Polytechnique F\'ed\'erale de Lausanne (EPFL), CH-1015 Lausanne, Switzerland}
\author{S. Gorno}
\affiliation{Swiss Plasma Center (SPC), \'Ecole Polytechnique F\'ed\'erale de Lausanne (EPFL), CH-1015 Lausanne, Switzerland}
\author{C. Theiler}
\affiliation{Swiss Plasma Center (SPC), \'Ecole Polytechnique F\'ed\'erale de Lausanne (EPFL), CH-1015 Lausanne, Switzerland}
\author{M. Carpita}
\affiliation{Swiss Plasma Center (SPC), \'Ecole Polytechnique F\'ed\'erale de Lausanne (EPFL), CH-1015 Lausanne, Switzerland}
\author{G. Durr-Legoupil-Nicoud}
\affiliation{Swiss Plasma Center (SPC), \'Ecole Polytechnique F\'ed\'erale de Lausanne (EPFL), CH-1015 Lausanne, Switzerland}
\author{M. von Allmen}
\affiliation{Swiss Plasma Center (SPC), \'Ecole Polytechnique F\'ed\'erale de Lausanne (EPFL), CH-1015 Lausanne, Switzerland}
\date{\today}

\begin{abstract}
Studying the process of divertor detachment and the associated complex interplay of plasma dynamics and atomic physics processes is of utmost importance for future fusion reactors. Whilst simplified analytical models exist to interpret the general features of detachment, they are limited in their predictive power, and complex 2D or even 3D codes are generally required to provide a self-consistent picture of the divertor. As an intermediate step, 1D models of the Scrape-Off Layer (SOL) can be particularly insightful as the dynamics are greatly simplified, while still self-consistently including various source and sink terms at play, as well as additional important effects such as flows. These codes can be used to shed light on the physics at play, to perform fast parameter scans, or to interpret experiments. In this paper, we introduce the SPLEND1D (\emph{Simulator of PLasma ENabling Detachment in 1D}) code: a fast and versatile 1D SOL model. We present in detail the model that is implemented in SPLEND1D. We then employ the code to explore various elements of detachment physics for parameters typical of the Tokamak à Configuration Variable (TCV), including the atomic physics and other processes behind power and momentum losses, and explore the various hypotheses and free parameters of the model. 

\end{abstract}
\maketitle 

\section{\label{sec:introduction}Introduction}
Plasma power and particle exhaust is a crucial issue for future fusion reactors. If unmitigated, the target heat fluxes in ITER and DEMO are expected to greatly exceed the $10~\mathrm{MW/m}^2$ limit that is considered necessary for tolerable steady-state conditions. It will thus be necessary to operate in an, at least partially, detached regime\cite{stangeby2000plasma,Krasheninnikov_PoP2016,Leonard_PPCF2018}. In such a regime, the total plasma pressure develops strong parallel gradients along the field lines in the \emph{Scrape-Off Layer} (SOL), driven by volumetric momentum losses \cite{Loarte_NF1998,Verhaegh_NF2018},  and a significant fraction of the plasma power is dissipated by volumetric power sinks. This results in a reduction of the target heat and particle flux densities, as well as target temperature, which is important to reduce target erosion \cite{Stangeby_NF2011,Kallenbach_PPCF2013}. The detachment process typically sets in at a target temperature below 5 eV \cite{Lipschultz_FST2007, Potzel_NF2014,Verhaegh_NF2018}, where plasma-neutral interactions are further enhanced, further reducing target heat flux, temperature, and ion flux.

This paper introduces the SPLEND1D code, a fast and versatile 1D model used to explore the complex interplay of atomic physics and plasma dynamics underlying the detachment process. The purpose of this paper is two-fold. First, to present in detail the model implemented in the SPLEND1D code. Second, to apply the code for the study of a base case scenario for parameters typical for the Tokamak à Configuration Variable (TCV) \cite{Reimerdes_NF2022}. The neutral particle source is increased in this scenario to reach the onset of detachment, similarly to the experimental onset of detachment through increased plasma fuelling, The different processes are subsequently investigated, thus providing a first illustration of the SPLEND1D capabilities and possible applications. This paper is organized as follows. In section \ref{sec:model}, the derivation of the models for the charged species (plasma) and the neutral species is presented, as well as the numerical methods, boundary conditions and implemented source terms. Section \ref{sec:basecase} presents a reference base case that is used to demonstrate the capabilities of the SPLEND1D code in terms of ease of use and result interpretation. We also present some measurements of the code accuracy and runtime. In section \ref{sec:detachment_basecase}, we simulate detachment of the base case through upstream density ramps, and highlight the mechanisms at play in the model that enable the onset of momentum and energy losses. Section \ref{sec:freeparam} investigates the role of some free parameters of the model, such as the impurity concentration, the neutral confinement time, the heat-flux limiters, and the choice of boundary condition for the parallel velocity. Finally, in section \ref{sec:change_model}, we present some advanced studies that have been enabled by SPLEND1D, investigating the choice of the neutral model, the effect of separating ion and electron energy equations, and the results of time-dependent simulations. These advanced studies demonstrate SPLEND1D's aptness in interpreting current TCV experiments, for example to study the role of connection length on divertor detachment \cite{Gorno2024}. Comparisons to other 1D codes in the community are discussed throughout the text.

\section{Model}\label{sec:model}
In this first section, we describe the model implemented in SPLEND1D. The equations solved by SPLEND1D are based on the Braginskii equations \cite{Braginskii_1965}, which are typically used to describe the evolution of the plasma in the SOL in 1D, 2D or 3D codes, using a fluid approximation. In particular, we present in detail the assumptions used to derive the SPLEND1D code.

\subsection{Geometry}
\begin{figure}
\includegraphics[width=\linewidth]{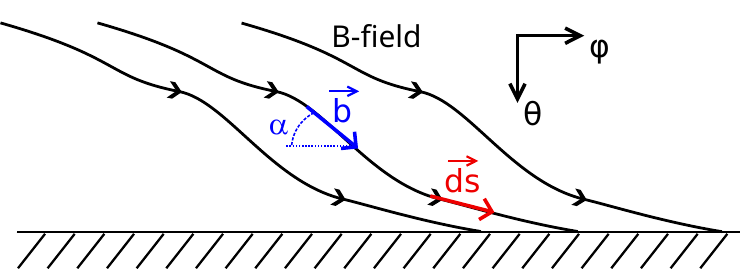}
\caption{\label{fig:drawing_Olivier_1D_detachment_model}\raggedright Definition of the geometry used in SPLEND1D. $\mathbf{b}$ is the unit vector parallel to the magnetic field. $\alpha$ corresponds to the pitch angle. $s$ is a curvilinear coordinate along the flux tube, oriented along $\mathbf{ds}$. $\theta$ and $\varphi$ are two spatial coordinates that will be relevant for the 2D neutral model that is presented in section \ref{subsec:advective_neutrals}.} 
\end{figure}

We consider a one-dimensional model for the SOL. The geometry of this 1D SOL is axisymmetric, i.e., uniform along $\varphi$ in Figure \ref{fig:drawing_Olivier_1D_detachment_model}, but otherwise arbitrary. In particular, both the magnitude ($B$) of the magnetic field ($\mathbf{B}$) as well as its pitch angle are allowed to vary along a field line. We introduce the curvilinear coordinate $s$ along the magnetic field, such that $\vec{\mathrm{d}s}$ is parallel to the unit vector $\mathbf{b} = \frac{\mathbf{B}}{B}$, Figure \ref{fig:drawing_Olivier_1D_detachment_model}. Here, $s$ is defined to increase towards the target, Figure \ref{fig:drawing_Olivier_1D_detachment_model}. We denote $\alpha(s)$ the local pitch angle. It can be related to the components of the magnetic field by
\begin{equation} 
\tan \alpha = \frac{B_\theta}{B_\varphi},
\end{equation}
where $B_{\theta}(s)$ and $B_\varphi(s)$ are the poloidal and toroidal components of the magnetic field, Figure \ref{fig:drawing_Olivier_1D_detachment_model}.

\subsection{Plasma model}\label{sec:plasma_model}
\subsubsection{Equations}\label{sec:plasma_eq}
In the following, we assume toroidal symmetry within the system, such that $\partial/\partial\varphi = 0$ for all considered quantities. The plasma fluid velocity is assumed purely parallel to the magnetic field, and to be the same for ions and electrons, such that there is no parallel current. The velocity vector $\mathbf{V}$ is written as $\mathbf{V} = u_\|\mathbf{b}$. The conductive and viscous heat fluxes are also assumed to be parallel to the magnetic field. We denote $n$ the plasma density, assuming quasi-neutrality ($n_i = n_e = n$ with $Z=1$), and $m_e$ (resp. $m_i$) the electron (resp. ion) mass. The neutral mass $m_n$ is taken equal to the ion mass, $m_n=m_i$. $T_i$ and $T_e$ are the ion and electron temperatures. We assume high enough collisionality, such that, for both ions and electrons, the pressure is isotropic. With these assumptions, we project the Braginskii equations along $\mathbf{b}$, which results in the following continuity equation, total parallel plasma momentum equation (obtained by summing up the electron and ion momentum equations assuming $m_e \ll m_i$), and electron and ion energy equations.
\begin{widetext}
\begin{align}
\frac{\partial n}{\partial t} + B\frac{\partial}{\partial s} \left( \frac{nu_\|}{B} \right) = S^n_p + H_P, \label{eq:continuity}\\
\frac{\partial}{\partial t}  \left(m_i n u_\|\right) + B\frac{\partial}{\partial s} \left( \frac{m_inu_\|^2}{B} \right) = -\frac{\partial}{\partial s}\left( p_e + p_i\right) + S^u_{p,\|}  - \mathbf{b}\cdot\nabla\cdot\mathbf{\Pi},\label{eq:df_momentum} \\
\frac{\partial}{\partial t} \left( \frac{3}{2} nT_e \right)+ B\frac{\partial}{\partial s} \left( \frac{\frac{5}{2}nT_eu_\| + q_{\|,e}^{cond}}{B} \right) = u_\| \frac{\partial p_e}{\partial s} + Q_e + S_e^E + S_{imp}^E + H_e,  \label{eq:eenergy}\\
\frac{\partial}{\partial t} \left( \frac{3}{2} nT_i + \frac{1}{2} m_i n u_\|^2 \right)+ B\frac{\partial}{\partial s} \left(\frac{\frac{5}{2}nT_iu_\| + \frac{1}{2}m_i n u^3_\| + q_{\|,i}^{cond}}{B} \right) = - u_\| \frac{\partial p_e}{\partial s} + Q_i + S_i^E + H_i  - \nabla\cdot\left(\mathbf{V}\cdot\mathbf{\Pi}\right).\label{eq:ienergy}
\end{align}
\end{widetext}
Here, quantities are defined in SI units. As for the temperatures, $T$ stands for ${k_B}T$, where ${k_B}$ is the Boltzmann constant. $p_e=n T_e$, $p_i = n T_i$ are the static pressures of the electrons and ions, respectively. $S^n_p$, $S^u_{p,\|} $, $S_e^E$ and $S_i^E$ are particle, momentum, electron energy and ion energy source terms resulting from ionization, recombination, charge-exchange and excitation reactions. They will be described in section \ref{sec:atomics}. The term $S_{imp}^E$ in equation (\ref{eq:eenergy}) corresponds to the energy loss due to impurity radiation, also described in section \ref{sec:atomics}. $q_{\|,e}^{cond}$ (resp. $q_{\|,i}^{cond}$) is the electron (resp. ion) parallel conductive heat flux, whose expressions will be given in section \ref{sec:heatfluxes}. $\mathbf{b}\cdot\nabla\cdot\mathbf{\Pi} $ and $\nabla\cdot\left(\mathbf{V}\cdot\mathbf{\Pi}\right)$ are the viscous contributions to the parallel momentum and energy equations, detailed in section \ref{sec:visc}. $Q_e$ and $Q_i$ model the exchange of energy between ions and electrons due to collisions, and can be expressed as
\begin{equation}
Q_e = -Q_i = 3\frac{m_e}{m_i}\frac{n}{\tau_e}\left(T_i - T_e\right),
\label{eq:equipartition}
\end{equation}
where $\tau_e$ is the electron collision time, defined as\cite{Braginskii_1965,NRL_Plasma_Formulary}
\begin{align}
\tau_e &= \frac{6\sqrt{2}\,\pi^{3/2}\,\epsilon_0^{~2}\,\sqrt{m_e}\,\,T_e^{~3/2}}{\ln\Lambda\, e^4\, n}, \label{eq:taue} \\
&\approx \frac{1}{3.64\times10^{-6}} \frac{\left(T_e\left[\mathrm{K}\right]\right)^{3/2}}{\ln\Lambda\, n\left[m^{-3}\right]} \\
&\approx \frac{1}{2.91\times10^{-12}} \frac{\left(T_e \left[\mathrm{eV}\right]\right)^{3/2}}{\ln\Lambda\, n\left[m^{-3}\right]}
\end{align} 
with $\ln \Lambda$ the Coulomb Logarithm. $H_P$ in equation (\ref{eq:continuity}) is an additional volumetric (charged) particle source term, and is an input for the code. It can be used to model a flux of particles entering or leaving the flux tube. Similarly, $H_e$ and $H_i$ in equations (\ref{eq:eenergy}) and (\ref{eq:ienergy}) are volumetric energy source terms. 

To reduce the number of degrees of freedom, the code can be run under the assumption $T_i = \bar{\tau} T_e$, with $\bar{\tau}>0$ an arbitrary constant. Summing equations (\ref{eq:eenergy}) and (\ref{eq:ienergy}), equations (\ref{eq:continuity})-(\ref{eq:ienergy}) can then be rewritten as
\begin{widetext}
\begin{align}
\frac{\partial n}{\partial t} &+ B\frac{\partial}{\partial s} \left( \frac{nu_\|}{B} \right) = S^n_p + H_P,\label{eq:sf_continuity} \\
\frac{\partial}{\partial t}  \left(m_i n u_\|\right) &+ B\frac{\partial}{\partial s} \left( \frac{m_inu_\|^2}{B} \right) = -\left(1+\bar{\tau}\right)\frac{\partial p_e}{\partial s}+ S^u_{p,\|} - \mathbf{b}\cdot\nabla\cdot\mathbf{\Pi}, \label{eq:sf_momentum} \\
\frac{\partial}{\partial t} \left( \frac{3}{2} n\left(1+\bar{\tau} \right)T_e + \frac{1}{2} m_i n u_\|^2 \right) &+ B\frac{\partial}{\partial s} \left(\frac{\frac{5}{2}n\left(1+\bar{\tau}\right)T_eu_\| + \frac{1}{2}m_i n u^3_\| + q_{\|,i}^{cond} + q_{\|,e}^{cond} }{B}\right) \nonumber \\ &= S_i^E + S_e^E + S_{imp}^E + H_e + H_i - \nabla\cdot\left(\mathbf{V}\cdot\mathbf{\Pi}\right),  \label{eq:sf_energy}
\end{align}
\end{widetext}
SPLEND1D is able to solve either the two fluids model (equations (\ref{eq:continuity})-(\ref{eq:ienergy})) or the one-fluid model (equations (\ref{eq:sf_continuity})-(\ref{eq:sf_energy})), depending on the user inputs.

\subsubsection{\label{sec:heatfluxes}Heat fluxes}
The electron and ion parallel conductive heat fluxes are defined as 
\begin{equation}
\frac{1}{q_{\|,e}^{cond}} = \frac{1}{q_{\|,e}^{lim}} + \frac{1}{q_{\|,e}^{SH}},
\end{equation}
and
\begin{equation}
\frac{1}{q_{\|,i}^{cond}} = \frac{1}{q_{\|,i}^{lim}} + \frac{1}{q_{\|,i}^{SH}},
\end{equation}
where $q_{\|,e}^{SH}$ and $q_{\|,i}^{SH}$ are defined by
\begin{equation}
q_{\|,e}^{SH} = - \kappa_{e}^{SH}\frac{\partial T_e}{\partial s}, \quad q_{\|,i}^{SH}= - \kappa_{i}^{SH}\frac{\partial T_i}{\partial s},
\end{equation}
where $\kappa_{e}^{SH}$ and $\kappa_{i}^{SH}$, the classical Spitzer-H{\"a}rm electron and ion heat conduction coefficients, are defined as\cite{Braginskii_1965}
\begin{equation}
     \kappa_{e}^{SH} = 3.16\frac{n\tau_e T_e}{m_e},
  \label{eq:spitzer_heatcond_electron}
\end{equation}
and 
\begin{equation}
     \kappa_{i}^{SH} = 3.9\frac{n\tau_i T_i }{m_i}, 
  \label{eq:spitzer_heatcond_ion}
\end{equation}
Here, $\tau_e$ is the electron collision time defined in equation (\ref{eq:taue}) and $\tau_i$ is the ion collision time defined as\cite{Braginskii_1965,NRL_Plasma_Formulary}
\begin{align}
\tau_i &= \frac{ 12\,\pi^{3/2}\,\epsilon_0^{~2}\,\sqrt{m_i}\,\,T_i^{~3/2}}{\ln\Lambda\, e^4\, n},\\
&\approx \frac{1}{6.0\times10^{-8}} \sqrt{\frac{m_i}{m_p}}  \frac{\left(T_e\left[\mathrm{K}\right]\right)^{3/2}}{\ln\Lambda\, n \left[m^{-3}\right]}, \\
&\approx \frac{1}{4.8\times10^{-14}} \sqrt{\frac{m_i}{m_p}} \frac{\left(T_e \left[\mathrm{eV}\right]\right)^{3/2}}{\ln\Lambda\, n \left[m^{-3}\right]},
\end{align} 
where $m_p$ is the proton mass and where we assumed $Z=1$. 
In low collisionality (long mean free path) regimes, however, the physical heat fluxes may be significantly overestimated by the classical Spitzer-H{\"a}rm heat fluxes \cite{Day_CPP1996,stangeby2000plasma,Fundamenski_PPCF2005,Ciraolo_CPP2018}. In order to avoid this non-physical divergence, we use so-called ``flux limiters'', that limit the maximum value of the heat flux to the free streaming heat flux $ q_{\|,\{i,e\}}^{lim}$, defined as 
\begin{equation}
q_{\|,{\{i,e\}}}^{lim} = \alpha_{\{i,e\}} n T_{\{i,e\}} v^{th}_{\{i,e\}} = \alpha_{\{i,e\}} n T_{\{i,e\}} \sqrt{\frac{T_{i,e}}{m_{i,e}}}.
\end{equation}
$\alpha_{i}$ and $\alpha_{e}$ are two free parameters, whose typical values are around 0.5 (Ref. \cite{Day_CPP1996,Fundamenski_PPCF2005}), although formally one would require kinetic simulations to determine the values of these parameters. By affecting the heat flux, they may influence the predictions of the simulation \cite{Day_CPP1996,Fundamenski_PPCF2005,Schneider_CPP2006}. In section \ref{sec:flux_limiter}, we will investigate the sensitivity of the results presented in this paper on these values.

\subsubsection{\label{sec:visc}Viscosity tensor}
SPLEND1D includes the effect of parallel viscosity. The term $\mathbf{b}\cdot\nabla\cdot\mathbf{\Pi}$, that appears in equations (\ref{eq:df_momentum}) and (\ref{eq:sf_momentum}), can be written as\cite{Braginskii_1965} 
\begin{equation}
\mathbf{b}\cdot\nabla\cdot\mathbf{\Pi} = \mathbf{b}\cdot\nabla\cdot \left[\delta p \left(\mathbf{bb}  - \frac{1}{3}\mathbb{1} \right)\right] = \frac{2}{3}\left[ B^{\frac{3}{2}}\frac{\partial}{\partial s} \left( \delta p B^{-\frac{3}{2}}\right) \right],
\end{equation}
where
\begin{equation}
\delta p = -\eta_\|\left( 3\frac{\partial u_\|}{\partial s} - \nabla\cdot u\right),
\end{equation}
and $\eta_\|$ is the parallel ion viscosity, defined later in equation (\ref{eq:par_visc}). This results in
\begin{equation}
\mathbf{b}\cdot\nabla\cdot\mathbf{\Pi} =  -\frac{4}{3} \left[ B^{\frac{3}{2}}\frac{\partial}{\partial s} \left( \eta_\| B^{-2}\frac{\partial B^{\frac{1}{2}}u_\|}{\partial s} \right) \right].
\label{eq:visc}
\end{equation}
The viscosity also contributes to the ion energy equation. Starting from the contribution of the viscosity to heat generation, $\nabla\cdot\left(u_\|\cdot \left[ \delta p \left(\mathbf{bb}  - \frac{1}{3}\mathbb{1} \right)  \right]\right)$, one finds an energy source term of the form
\begin{equation}
\nabla\cdot\left(\mathbf{V}\cdot\mathbf{\Pi}\right)  = -\frac{4}{3} \left[ B\frac{\partial}{\partial s} \left( u_\|\eta_\|B^{-\frac{3}{2}}\frac{\partial B^{\frac{1}{2}}u_\|}{\partial s} \right) \right].
\label{eq:visc_diss}
\end{equation}
Similarly to the classical heat fluxes, the viscous flux can become unphysical at low collisionality, and hence requires a flux limiter. This is done by writing
\begin{equation}
{\eta_\|} = \frac{\eta_\|^{br}}{1 + \frac{\eta_\|^{br}\left| \frac{\partial u_\|}{\partial s}\right|}{\tau_\|^{lim} }} + n\nu_{num},
\label{eq:par_visc}
\end{equation}
where $\tau_\|^{lim} = \frac{4}{7} n T_i$, and $\eta_\|^{br}$ is the Braginskii ion parallel viscosity, expressed as $\eta_\|^{br} = 0.96n T_i \tau_i$. More complete expressions of this flux limiter can be developed \cite{Zawaideh_PoF1986, Zawaideh_PoF1988,Fundamenski_PPCF2005, Havlickova_PPCF2013}, although in this paper we restrict ourselves to this simple form, which is similar to the one implemented in 2D transport codes such as SOLPS-ITER \cite{Schneider_CPP2006, Havlickova_PPCF2013}. $\nu_{num}$ is a numerical kinematic viscosity that can be employed to facilitate numerical convergence, but that is not enabled by default in the code.

\subsubsection{\label{sec:coulomb_logarithm}Coulomb logarithm}
The Coulomb Logarithm, $\ln \Lambda$, is a slow varying function of density and temperature, and is therefore typically set constant. It is, however, also possible to enforce a local computation of $\ln \Lambda$, based on the local $n_e$ and $T_e$. The usual definition of $\ln \Lambda$ yields\cite{NRL_Plasma_Formulary}, for $n_e$ expressed in $\mathrm{m^{-3}}$ and $T_e$ in $\mathrm{eV}$,
\begin{equation}
\ln \Lambda = \ln \Lambda_{low} = 23-\mathrm{ln}\left( \left(n_e\times10^{-6}\right)^{0.5} T_e^{-1.5}\right),
\end{equation}
for $T_e < 10~\mathrm{eV}$ and 
\begin{equation}
\ln \Lambda = \ln \Lambda_{high} = 24-\mathrm{ln}\left( \left(n_e\times10^{-6}\right)^{0.5} T_e^{-1}\right),
\end{equation}
for $T_e > 10~\mathrm{eV}$. Using such a definition of $\ln \Lambda$ leads to a discontinuity at $T_e=10~\mathrm{eV}$, that causes numerical difficulties to the non-linear solver by introducing a singularity in the Jacobian. To avoid this issue, we introduce a transition parameter, $\Delta$, defined as
\begin{equation}
\Delta = \frac{1}{2}\left[\mathrm{tanh}\left( \frac{T_e~\mathrm{[eV]}-10~\mathrm{[eV]}}{0.1~\mathrm{[eV]}}\right)+1\right],
\end{equation}
such that 
\begin{equation}
\ln \Lambda = (1-\Delta) \ln \Lambda_{low} + \Delta \ln \Lambda_{high}.
\label{eq:coulomb_logarithm_smooth}
\end{equation}
Equation (\ref{eq:coulomb_logarithm_smooth}) is a smooth function of $n_e$ and $T_e$, thus avoiding the numerical difficulties associated with a discontinuous $\ln \Lambda$.

\clearpage
\subsection{Fluid model for the neutrals}
This section presents the model retained to describe the dynamics of the neutrals. As the neutrals are not affected by the magnetic field, their dynamics are intrinsically 3D. Furthermore, in typical divertor conditions, the mean free path of the neutrals can be large compared to the system size, leading to a high Knudsen number that in principle requires a kinetic description of the neutral dynamics. This is the approach of EIRENE, one of the main workhorse for neutral dynamics simulations in the fusion community\cite{Reiter_FST2005}, where a Monte Carlo method is used to simulate the behavior of the neutrals. However, such methods are computationally expensive and subject to statistical noise. In recent years, to alleviate these costs, there has been significant work devoted to the development of advanced fluid neutral models, or hybrid neutral models, mixing a kinetic and a fluid description, Ref \cite{Horsten_NME2017,Horsten_NME2022,VanUytven_NF2022} and references therein. As 1D models aspire for simplicity, in this work, we use fluid neutral models. We only model a single population of neutral atoms (no molecules). SPLEND1D implements two different neutral models, the choice of which to use is made by the user: a diffusive neutral model, in which the neutrals diffuse along the $\theta$-direction; and an advective one, where neutrals move in the $\theta-\phi$ plane (that is, a flux surface), with a velocity that is not necessarily parallel to the magnetic field, similar to the model presented in Refs. \cite{Horsten_CPP2016,Horsten_PSI2016}. 
In the following, we further assume that the vectors describing the neutral dynamics (velocities, heat fluxes) do not have a component in the radial direction ($\Psi$). Therefore, while neutrals are not confined to a particular flux tube, they remain confined to a particular flux-surface.

\subsubsection{Continuity equation for the neutrals}
The general form for the continuity equation for the neutral particles is given by 
\begin{equation}
\frac{\partial n_n}{\partial t}  + \nabla\cdot\left(n_n\mathbf{V}_n\right) = S^n_n  + H_n - \frac{n_n}{\tau_n},
\label{eq:continuity_neutrals}
\end{equation}
where $\mathbf{V}_n$ is the neutral velocity, $S^n_n$ is a neutral particle source (or sink) term due to atomic processes, that will be discussed in section \ref{sec:atomics}, and $H_n$ is an arbitrary neutral source or sink, set as input of the code, for instance to simulate fuelling of the plasma by neutrals, or simulate a neutral background, as done in Ref. \cite{Derks_PPCF2022}. Since neutrals are not bound to the magnetic field, they may escape the flux tube. This is modelled by an ad-hoc sink term \cite{Togo_PFR2013, Dudson_PPCF2019}, characterized by the characteristic neutral retention time $\tau_N$. {In the $(\theta,\Psi,\phi)$ coordinate system, $-\frac{n_n}{\tau_n}$ would include the $\Psi$ contribution to $\nabla\cdot\left(n_n\mathbf{V}_n\right)$, such that $-\frac{n_n}{\tau_n} = -\frac{\partial}{\partial \Psi}\left( n_n V_n^\Psi\right)$.} 

We now write $\mathbf{V}_n$ in terms of components, $\mathbf{V}_n(s) = (V_n^\theta(s), V_n^\varphi(s))$. After developing the divergence, equation (\ref{eq:continuity_neutrals}) can then be rewritten as
\begin{equation}
\frac{\partial n_n}{\partial t} + B\frac{\partial}{\partial s} \left( \frac{n_nV_n^\theta}{B\sin\alpha} \right) = S^n_n + H_n - \frac{n_n}{\tau_n}.
\label{eq:neutral_dens_} 
\end{equation}
To solve this equation, a description of $\mathbf{V}_n(s) = (V_n^\theta(s), V_n^\varphi(s))$ is required. This is presented in the subsequent sections. 

\subsubsection{Diffusive neutral model}
In the diffusive neutral model, we assume that the neutral flux can be written as 
\begin{equation}
n_n\mathbf{V}_n = -D_n \nabla n_n, 
\end{equation}
leading to 
\begin{align}
		V_n^\theta &= -\frac{D_n}{n_n\sin\alpha} \frac{\partial n_n}{\partial s}, \label{eq:vn_t_diff}\\
		V_n^\phi &= 0 \label{eq:vn_p_diff},
\end{align}
where $D_n$ is a diffusion coefficient, defined as
\begin{equation}
D_n = \frac{T_n}{m_n\left({n\left<\sigma v\right>_{cx}}+{n\left<\sigma v\right>_{ion}} \right )},\label{eq:dn}
\end{equation}
where $T_n$ is the neutral temperature, assumed, for this diffusive neutral model, to either match the ion temperature, $T_n=T_i$, or to be a constant (with a value specified by the user). $\left<\sigma v\right>_{cx}$ is the local charge-exchange reaction rate and $\left<\sigma v\right>_{ion}$ the local ionization reaction rate, that will be introduced in section \ref{sec:rates}. We remark here that ion-neutral elastic collisions are not considered, as we assume that the neutral and ion populations interact only through atomic processes (ionization, recombination, charge-exchange). We then obtain a diffusion equation for the neutral density $n_n$, 
\begin{equation}
\frac{\partial n_n}{\partial t} = B\frac{\partial }{\partial s}\left( \frac{ D_n}{B\sin^2\alpha}\frac{\partial n_n}{\partial s}\right) + S_n^n + H_n^n - \frac{n_n}{\tau_n}. \label{eq:nn_diff}
\end{equation}
This neutral model is similar to the one implemented in \cite{Nakazawa_PPCF2000, Derks_PPCF2022}, although we retain here a dependence on $T_n$, whereas Ref\cite{Nakazawa_PPCF2000} assumes $T_n = T_e$ and $B$ constant along the flux tube. 

\subsubsection{Advective neutral model}\label{subsec:advective_neutrals}
In the case of the advective neutral model, the $n_n\mathbf{V}_n$ term is described by the addition of a momentum equation. We also add ad-hoc loss terms involving $\tau_n$ in the momentum and energy equations, that will be developed hereafter. We start from the general form of the momentum equation
\begin{align}
\frac{\partial \left(m_n n_n\mathbf{V}_n\right)}{\partial t}  + \nabla\cdot\left(m_n n_n\mathbf{V}_n\otimes\mathbf{V}_n\right) &= -\nabla p_n + \mathbf{S^u_n} \\
&- \frac{m_n n_n\mathbf{V}_n}{\tau_n} - \nabla\cdot\mathbf{\Pi_n},
\end{align}
where $p_n = n_n T_n$ is the static neutral pressure, $\mathbf{S^u_n}$ is the neutral momentum source term, and $\mathbf{\Pi_n}$ the viscous stress tensor. In principle, the viscous stress tensor could be self-consistently evaluated. However, due to the complexity of the resulting expression (see Ref \cite{Horsten_CPP2016} for instance), we use a simpler form. Projected along the $\theta$ and $\varphi$ directions, and using $\frac{\partial}{\partial \varphi} = 0$, we get
\begin{widetext}
\begin{align}
\frac{\partial \left(m_nn_nV_n^\theta\right)}{\partial t} + B\frac{\partial}{\partial s} \left( \frac{n_n m_n {V_n^\theta}^2}{B\sin\alpha} \right) + \frac{n_n m_n{V_n^\varphi}^2}{B_\varphi\sin\alpha}\frac{\partial B_\varphi}{\partial s} &= -\frac{1}{\sin\alpha}\frac{\partial p_n}{\partial s} +{S^u_{n,\theta}} - \frac{m_nn_nV_n^\theta}{\tau_n} +\frac{\partial}{\partial s} \left( \frac{\eta_n }{\sin^2 \alpha} \frac{\partial V_n^\theta}{\partial s}  \right),\label{eq:neutral_vt} \\
\frac{\partial \left(m_nn_nV_n^\varphi\right)}{\partial t} + B\frac{\partial}{\partial s} \left( \frac{n_n m_n {V_n^\theta}{V_n^\varphi}}{B\sin\alpha} \right) - \frac{n_n m_n{V_n^\theta}{V_n^\varphi}}{B_\varphi\sin\alpha}\frac{\partial B_\varphi}{\partial s} &= {S^u_{n,\phi}} - \frac{m_nn_nV_n^\varphi}{\tau_n} +\frac{\partial}{\partial s} \left( \frac{\eta_n }{\sin^2 \alpha} \frac{\partial V_n^\varphi}{\partial s}  \right). \label{eq:neutral_vp}
\end{align}
\end{widetext}
where the neutral viscosity $\eta_n$ is defined as
\begin{equation}
\eta_n = \frac{n_n T_n}{\left({n\left<\sigma v\right>_{cx}}+{n\left<\sigma v\right>_{ion}} \right )},
\label{eq:eta_n}
\end{equation}

This model allows for neutral trajectories that are not aligned to the magnetic field, but constrained within a flux surface. For the neutral energy equation, the procedure is similar. Starting from the general form of the equation, we have
\begin{equation}
\frac{\partial E_n}{\partial t}  + \nabla\cdot\left( \left[E_n+p_n\right]\mathbf{V}_n + \mathbf{q}_n + \mathbf{V}_n \cdot\left(\nabla\cdot\mathbf{\Pi_n}\right) \right) = S^E_n - \frac{E_n}{\tau_N},
\end{equation}
where $S^E_n$ represents the source and sink terms, discussed in section \ref{sec:atomics}. $E_n$ is the total energy, defined as
\begin{equation}
E_n = \frac{3}{2} n_nT_n + \frac{1}{2} n_n m_n  V_n^2,
\end{equation}
where $V_n^2 = {V_n^\theta}^2 + {V_n^\varphi}^2$. $\mathbf{q}_n$ is the neutral conductive heat flux. Denoting $\kappa_n$ the neutral heat conduction, we write \cite{VanUytven_NF2022}
\begin{equation}
\kappa_n = \frac{5}{2} \frac{\eta_n}{m_n} = \frac{5}{2} \frac{n_n T_n}{m_n\left({n\left<\sigma v\right>_{cx}}+{n\left<\sigma v\right>_{ion}} \right )},
\end{equation}
following similar notations as for equation (\ref{eq:eta_n}). The neutral heat flux $\mathbf{q}_n$ is then defined as
\begin{equation}
\mathbf{q}_n = -\kappa_n \nabla T_n.
\end{equation}

Developing the divergence, this leads to
\begin{widetext}
\begin{equation}
\frac{\partial E_n}{\partial t} + B\frac{\partial}{\partial s} \left( \frac{\left[E+p_n\right] V_n^\theta + q_n^\theta + \left[\mathbf{V}_n \cdot\left(\nabla\cdot\mathbf{\Pi_n}\right)\right]_\theta }{B\sin\alpha} \right) = S^E_n - \frac{E_n}{\tau_n}.
\label{eq:neutral_energy}
\end{equation}
\end{widetext}
Equations (\ref{eq:neutral_dens_}), (\ref{eq:neutral_vt}), (\ref{eq:neutral_vp}) and (\ref{eq:neutral_energy}) constitute a set of four equations that are solved to describe the behavior of the neutrals. In Appendix \ref{app:pressure_diffusion}, we briefly show how equations (\ref{eq:neutral_dens_}), (\ref{eq:neutral_vt}), (\ref{eq:neutral_vp}) could be further developed to form a so-called ``pressure-diffusion" model, in the spirit of models that have been developed in Ref\cite{Horsten_NME2017}. The implementation and test of such formulation in SPLEND1D is, however, left for future work. 

\subsection{Atomic source terms}\label{sec:atomics}
Now, we introduce the source terms resulting from atomic interactions between the neutral and plasma populations. We restrict ourselves here to the atomic processes of ionization, charge exchange, excitation, and recombination since we model a single-ion population and consider only atomic neutrals. We remark here that ion-neutral elastic collisions are not considered, as we assume that the neutral and ion populations interact only through these atomic processes (ionization, recombination, charge-exchange). Inclusion of ion-neutral elastic collisions is left for future work. 

\subsubsection{Rates}\label{sec:rates}
SPLEND1D's rate coefficients are either obtained from a bi-linear interpolation of the open-ADAS\cite{Summers_PPCF2006} database, or evaluated using the fit provided in the AMJUEL manual \cite{AMJUEL}, as decided by the user. Table \ref{tab:adas_amjuel_coefficient} presents the sources used for the ionization, excitation, recombination and charge exchange rates as functions of $n_e$ and $T_e$. The charge-exchange reaction is discussed further below. In the case of AMJUEL data, the excitation rate $Q_{exc}$ includes both the power radiated by excitation and the potential energy cost $E_{ion}$ in case of ionization ($E_{ion} = 13.6~\mathrm{eV}$ in the case of hydrogen)\cite{AMJUEL}. In the case of open-ADAS data, the latter is added through an additional sink term proportional to $E_{ion}$ and the ionization rate.

\begin{table*}[ht]
\begin{center}
\begin{tabular}{| c | c | c | c |}
\hline
Reaction & Rate coefficient & AMJUEL \cite{AMJUEL}&  open-ADAS\cite{Summers_PPCF2006} \\
\hline
Ionization& $\left< \sigma v\right>_{ion}$ & Reaction 2.1.5, section 4 & \texttt{SCD} \\
Excitation& $Q_{exc}$ & Reaction 2.1.5, section 10.2 & \texttt{PLT}\\
Recombination& $\left< \sigma v\right>_{rec}$ & 2.1.8, section 4 & \texttt{ACD} \\
Recombination cooling rate& $Q_{rec}$ & 2.1.8, section 10.4 & \texttt{PRB} \\
Charge exchange& $\left< \sigma v\right>_{CX}$ & 2.1.9, section 3.1.8* & \texttt{CCD} \\
\hline
\end{tabular}
\caption{{\label{tab:adas_amjuel_coefficient}\raggedright Summary of the rate coefficients used in SPLEND1D. Note: The charge-exchange reaction is discussed further in section \ref{sec:rates}}}
\end{center}
\end{table*}

The charge-exchange rate coefficient $\left< \sigma v\right>_{CX}$, when evaluated using the AMJUEL fit, is a function of an effective temperature defined, following the AMJUEL reference manual, as
\begin{equation}
T_{eff} = \frac{m_{i}}{m_H}\left(T_i + \beta T_n\right),
\label{eq:teff}
\end{equation}
where $m_{i}$ is the ion mass used in SPLEND1D, ${m_H}$ the hydrogen (proton) mass, and $\beta$ is a user flag, set equal to either 0 or 1, determining whether or not to include the finite neutral temperature in the calculation of the effective temperature. Note that in equation (\ref{eq:teff}), we used the assumption $m_n = m_i$. In the case of the open-ADAS \texttt{CCD} coefficient, no rescaling of $T_{eff}$ is performed, and the rate is computed from $T_e$. It is then the responsibility of the user to ensure that the provided rate is indeed adequate. 

Since these rates are estimated from either tabulated data (open-ADAS) or from fitted expressions, they are only defined over a certain range of validity. If, during a SPLEND1D simulation, the temperature or density were to go outside of these ranges, the rates are computed using the minimum (or maximum) values for which they are defined.  

\subsubsection{Sources and sinks}
The plasma particle source resulting from these atomic reactions is given by
\begin{equation}
S^n_p = \underbrace{nn_n\left< \sigma v\right>_{ion}}_{ionization} \underbrace{-n^2\left< \sigma v\right>_{rec}}_{recombination}.
\end{equation}
For the neutrals, we have $S^n_n = -S^n_p$. 

For the energy equation, the source term is separated between the electron and ion contributions. For the electrons, one has \cite{Havlickova_PPCF2013,Horsten_PSI2016}
\begin{align}
S_e^E &=  \underline{\underbrace{{-nn_n\left< \sigma v\right>_{ion}E_{ion}}}_{ionization}} \nonumber \\ &  \underbrace{-nn_n Q_{exc}}_{excitation}   \underbrace{+n^2(E_{ion}\left< \sigma v\right>_{rec} - Q_{rec})}_{recombination}.
\label{eq:atomics_electrons_energy}
\end{align}
The \textit{ionization} term in equation (\ref{eq:atomics_electrons_energy}) is added only when using open-ADAS coefficients. As mentioned above, it is already included in the $Q_{exc}$ terms derived from AMJUEL. 
{At this point, it is important to justify the recombination contribution to the electron energy balance, equation (\ref{eq:atomics_electrons_energy}). The term is split between two contributions, 
$n^2E_{ion}\left< \sigma v\right>_{rec}$, that releases the ionization potential energy back to the electrons, and the $-n^2Q_{rec}$ term, which encompasses the radiative energy losses during recombination, as well as some further Bremsstrahlung losses if $Q_{rec}$ is taken from the open-ADAS \texttt{PRB} coefficient\cite{Summers_PPCF2006}. This is discussed in Ref. \cite{stangeby2000plasma}, chapter 3, and in Ref. \cite{Verhaegh_NF2021}, section 4.3. In particular, it was found that in typical TCV conditions, these two contributions approximately balance each other \cite{Verhaegh_NF2018}. This may, however, not be the case when three-body recombination dominates. We remark here that this implementation is consistent with that of EMC3-EIRENE \cite{Frerichs_PoP2021} and SolEdge-2D (fluid neutrals) \cite{Valentinuzzi_these2018}. We further remark that this term can disabled by the user in SPLEND1D. It will however be included in the simulations presented in this paper, although its contribution to the energy balance of the simulations is marginal.}

For the ions, we have
\begin{align}
S_i^E & = \underbrace{nn_n\left< \sigma v\right>_{ion} \left[ \frac{3}{2}T_n + \frac{1}{2}m_n  {V_n}^2 \right]}_{ionization} \nonumber \\ 
          & \underbrace{-  n^2\left< \sigma v\right>_{rec}\left[\frac{3}{2}T_i + \frac{1}{2}m_i u_{\|}^2 \right]}_{recombination}  \nonumber \\
          & \underbrace{ + n_n n \left< \sigma v\right>_{CX}\left[ \frac{3}{2}\left( T_n - T_i\right) + \frac{1}{2}\left( m_n V_n^2 - m_i u_{\|}^2\right) \right]}_{charge-exchange}.
\label{eq:atomics_ions_energy}
\end{align} 
For the neutral energy equation, we have $S_n^E = - S_i^E$, so that the energy that is lost, or gained, by the ions during atomic reactions is transferred to, or from, the neutrals.

Regarding the source terms in the plasma momentum equation, we have
\begin{align}
\mathbf{S^u_p} &= \underbrace{m_inn_n\left< \sigma v\right>_{ion}\mathbf{V_n}}_{ionization} \nonumber \\& \underbrace{-m_in^2\left< \sigma v\right>_{rec}\mathbf{V}}_{recombination} \nonumber \\& \underbrace{+m_in_n n \left< \sigma v\right>_{CX} \left( \mathbf{V_n} - \mathbf{V} \right)}_{charge-exchange}.
\label{eq:momentum_equation_vector}
\end{align}
We project equation (\ref{eq:momentum_equation_vector}) along $\mathbf{b}= ( \sin\alpha, \cos\alpha)$, and find
\begin{align}
{S^u_{p,\|}} &= \underbrace{m_inn_n\left< \sigma v\right>_{ion}\left(\sin\alpha  {V_n^\theta} + \cos\alpha  {V_n^\phi}\right)}_{ionization} \nonumber \\& \underbrace{-m_in^2\left< \sigma v\right>_{rec}u_\|}_{recombination} \nonumber \\& \underbrace{+m_in_n n \left< \sigma v\right>_{CX} \left( \left(\sin\alpha  {V_n^\theta} + \cos\alpha  {V_n^\phi}\right) -  u_\| \right)}_{charge-exchange}. \label{eq:momentum_term_expression}
\end{align}
This implies that, within our model, when a neutral ionizes, the resulting ion inherits only the parallel component of the neutral momentum. The perpendicular component of the neutral momentum is lost, leading to an effective increase in the resulting ion thermal energy, because the parallel momentum of the neutral is transferred to the ion, as well as its total energy. Therefore, the perpendicular energy $\frac{1}{2}m_n n_n \left(V_n^\perp\right)^2$ is redistributed as thermal energy. 

For the neutral momentum source terms, we project $\mathbf{S^u_n} = -\mathbf{S^u_p}$ onto the ($\theta, \varphi$) basis. This yields 
\begin{widetext}
\begin{equation}
\begin{cases}
	 {S^u_{n,\theta}} = \underbrace{-m_inn_n\left< \sigma v\right>_{ion}{V_n^\theta}}_{ionization}   \underbrace{+ m_in^2\left< \sigma v\right>_{rec}u_{\|}\sin\alpha}_{recombination}  \underbrace{- m_in_n n \left< \sigma v\right>_{CX}\left( {V_n^\theta} - u_{\|}\sin\alpha  \right)}_{charge-exchange},\\
	 {S^u_{n,\varphi}}  = \underbrace{-m_inn_n\left< \sigma v\right>_{ion}{V_n^\varphi}}_{ionization}  \underbrace{+ m_in^2\left< \sigma v\right>_{rec}u_{\|}\cos\alpha}_{recombination} \underbrace{- m_in_n n \left< \sigma v\right>_{CX}\left( {V_n^\varphi} - u_{\|}\cos\alpha  \right)}_{charge-exchange}.
\end{cases}
\end{equation}
\end{widetext}
The charge exchange reaction introduces a friction term that tends to align the neutral velocity with the magnetic field. In the case of the advective neutral model, it acts as a sink for their perpendicular momentum. 

Impurities are considered to only contribute to volumetric power loss, where all loss mechanisms are grouped into a single term, $S_{imp}^E$. We do not consider any dilution of the main plasma ions by the impurity species. For simplicity, we assume that for a given impurity ${imp}$, one has $n_{imp} = f_{imp} n$ where $f_{imp}$ is the impurity fraction and $n_{imp}$ the impurity density. $f_{imp}$ is assumed to be constant over the flux tube, and therefore we do not self-consistently model the distribution of the impurity density. Further, we assume coronal equilibrium, and the power loss due the impurities is written as
\begin{equation}
S_{imp}^E = - \sum_{imp} f_{imp} n^2 L_{z,{imp}}\left(T_e\right), 
\end{equation}
where $L_{z,{imp}}$ is the cooling-rate of the impurity ${imp}$, either taken from Ref. \cite{Post_ADNDT77} or pre-computed using a collisional radiative model (CRM)\cite{Wagner_2016} that employs open-ADAS. In this paper, we use the latter. If the electron temperature is outside of the range considered in these two sources, the electron temperature used in the evaluation of $L_{z,{imp}}\left(T_e\right)$ is clamped either to the minimum or maximum value of $T_e$ over which $L_{z,{imp}}\left(T_e\right)$ is defined. 

\subsection{Boundary conditions}
\begin{figure}
\centering
\includegraphics[width=1.00\linewidth]{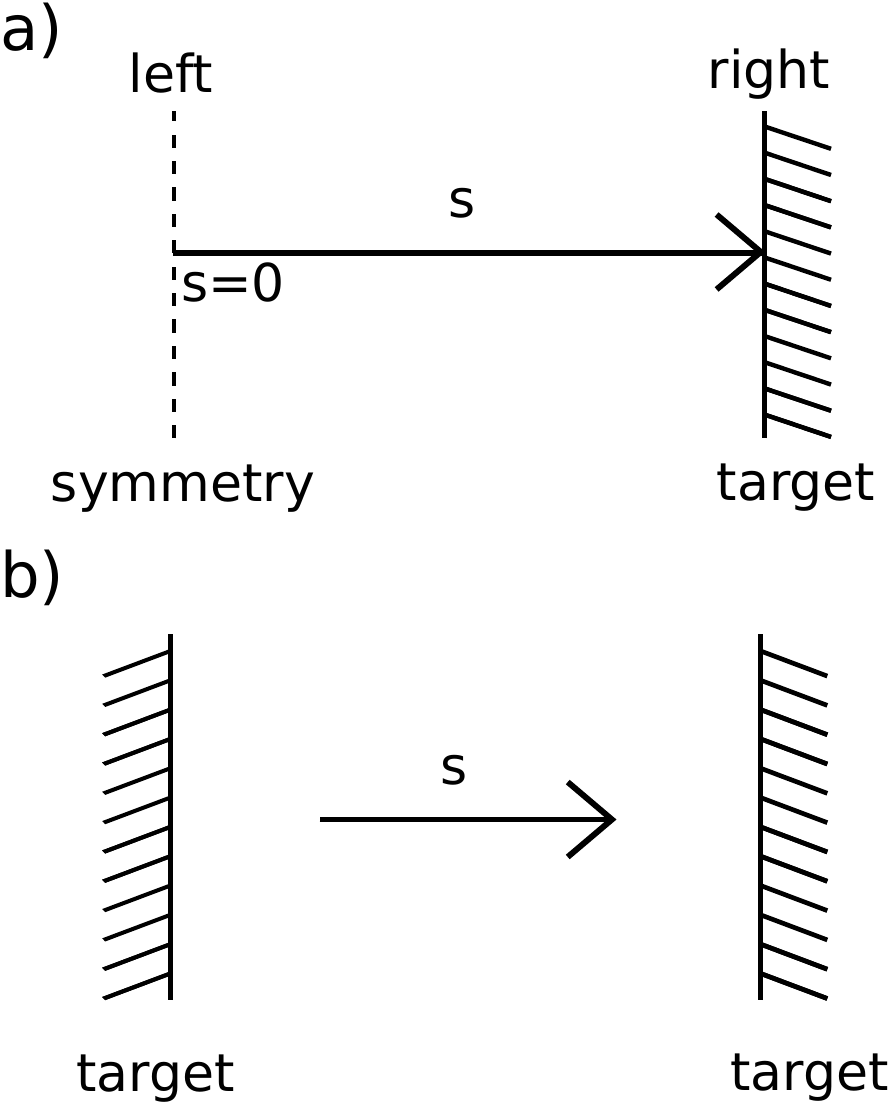}
\caption{\label{fig:boundary_fig}\raggedright Geometries simulated by SPLEND1D, drawn here for the situation $\alpha = \pi/2$. In the first case (panel a), the flux-tube is assumed to be symmetric around $s=0$. Symmetric boundary conditions are applied at the left boundary, while ``target'' boundary conditions are applied at the right boundary. In the second case (panel b), the flux tube connects two targets together, and there is not necessarily a symmetry plane. In this case, ``target'' boundary conditions are applied to the left and right boundaries. In both cases, $s$ is oriented so that it increases towards the right.}
\end{figure}

To solve the equations describing the evolution of the plasma parameters (equations (\ref{eq:continuity})-(\ref{eq:ienergy}) or (\ref{eq:sf_continuity})-(\ref{eq:sf_energy})) and the evolution of the fluid neutral parameters (equations (\ref{eq:neutral_dens_}), (\ref{eq:neutral_vt}), (\ref{eq:neutral_vp}) and (\ref{eq:neutral_energy})), a set of boundary conditions is required. We distinguish two types of boundary, depending on whether the considered boundary is a symmetry plane or a ``target''. Two kinds of configurations can be simulated with SPLEND1D (Figure \ref{fig:boundary_fig}): 
\begin{enumerate}[(a)]
\item The flux-tube is assumed to be symmetric around $s=0$. Symmetric boundary conditions are applied on the left boundary, while ``target'' boundary conditions are applied to the right boundary (Figure \ref{fig:boundary_fig}a).
\item The flux tube connects two targets together, and no symmetry is assumed. In this case, ``target'' boundary conditions are applied to the left and right boundaries (Figure \ref{fig:boundary_fig}b). 
\end{enumerate} 
The second case can be used, for instance, to study the power sharing between the two sides of the flux tube, which depends on the ratio of the connection lengths\cite{Maurizio_NME2019}.

In the case of a symmetry plane, at $s=0$, we impose the derivative of the scalar fields to be zero: \begin{equation}
\frac{\partial n}{\partial s} = \frac{\partial n_n}{\partial s} = 0, \frac{\partial T_e}{\partial s} = \frac{\partial T_i}{\partial s} = \frac{\partial T_n}{\partial s} = 0,
\end{equation}
while no-flow conditions are imposed for the velocities, such that
\begin{equation}
u_\| = 0, V_n^\theta = V_n^\varphi = 0.
\end{equation}

In the case of a ``target" boundary condition, we apply the Bohm criterion on the plasma velocity, imposing to to either match or exceed the sound speed $c_s$, as decided by the user. The sound speed is defined as the isothermal sound speed, that is, \cite{stangeby2000plasma}
\begin{equation}
c_s = \sqrt{\frac{\left(T_e + T_i \right)}{m_i}},
\label{eq:soundspeed}
\end{equation}
with the plus sign if the boundary is at the right side of the domain, and the minus sign if the boundary is at the left side. We remark here that expressing $c_s$ as the isothermal sound speed is an assumption of the model, as various expressions of $c_s$ can be used, depending on the sheath model retained (see Refs. \cite{stangeby2000plasma,Riemann_JoPD1991} for a discussion of this topic). The isothermal sound speed has been retained in SPLEND1D as this is the common choice in the community (for instance in the SOLPS-ITER code \cite{Wiesen_JNM2015}). 
In addition to constraining the velocity at the sheath entrance, we also enforce boundary conditions on the heat flux through the sheath, such that
\begin{equation}
	\begin{cases}
		q_{\|,t}^{i}  = \frac{5}{2}nT_iu_\| + \frac{1}{2}m_i n u^3_\| + q_{\|,i}^{cond} + q_{\|,i}^{visc} = \gamma_i nT_iu_\|,\\
		q_{\|,t}^{e} = \frac{5}{2}nT_eu_\|+ q_{\|,e}^{cond} = \gamma_e nT_eu_\|,
    \end{cases}
\end{equation}
where $q_{\|,t}^{i}$ and $q_{\|,t}^{e}$ are the target ion and electron heat fluxes, and $q_{\|,i}^{visc}$ is the viscous contribution to the heat flux. $\gamma_i$ and $\gamma_e$ are the sheath transmission coefficients, typically chosen as $\gamma_i=3.5$ and $\gamma_e=5.5$, see Chapter 2 of Ref. \cite{stangeby2000plasma}. {For simplicity, in this paper, we do not retain any dependency of $\gamma_i$ and $\gamma_e$ with plasma parameters, although they could in principle depend on local plasma parameters and the Mach number \cite{stangeby2000plasma}.} {We also further remark here that the boundary condition is applied on the total heat flux, which includes the viscous contribution $q_{\|,i}^{visc}$. In other codes, this term is often either neglected, or explicitly not included in the heat flux boundary condition. Testing the two formulations with SPLEND1D, we find that, in the conditions of the base case that will be reported in section \ref{sec:basecase}, such choice has negligible impact on the simulation outcomes, showing a modest effect only in strongly attached situations.}

Finally, for the neutrals, we impose a recycling boundary condition on the neutral flux $\mathbf{\Gamma_n} = n_n \mathbf{v_n}$, such that 
\begin{equation}
\mathbf{\Gamma_n}\cdot\mathbf{n} = -R \mathbf{\Gamma}\cdot\mathbf{n} 
\end{equation}
where $\mathbf{\Gamma}$ is the ion flux to the target, $R$ is the recycling coefficient, and $\mathbf{n}$ the normal vector to the target. In the case of the advective model, the boundary conditions for the neutral velocities and temperature are chosen as 
\begin{equation}
	\begin{cases}
		V_n^\theta   =  \mp\sqrt{\frac{E_0}{m_i}}, \\ 
		V_n^\varphi = 0, \\
		T_n = \frac{1}{3}{E_0}
    \end{cases}
\end{equation}
where $E_0$ is the Franck-Condon energy, taken as $E_0= 3~\mathrm{eV}$ \cite{Rognlien_FED2018}. In the case of the diffusive model, $V_n^\theta$ and $V_n^\varphi$ are set according to equations ($\ref{eq:vn_t_diff}$) and ($\ref{eq:vn_p_diff}$), respectively.

\subsection{Numerical implementation}\label{sec:num_details}
Implemented in Fortran, SPLEND1D employs a finite volume method, solving the plasma and neutrals equations in their conservative forms. The grid is typically chosen to be non-uniform and accumulate towards the boundaries of the computational domain where sheath boundary conditions are enforced, to account for the strong gradients that can develop there. For a symmetric case, the width $h_i$ of a given cell $i$ is typically defined as 
 \begin{equation}
 h_i = qh_{i-1},
 \label{eq:cellsize}
 \end{equation}
 where $0 < q \leq 1$. A third-order CWENO3 reconstruction \cite{Puppo_JSC2014}, associated with Rusanov numerical fluxes, is used for the advection terms. {A low-order reconstruction, where quantities are assumed constant in a cell, is also implemented. While this further speeds-up the code, this comes at the price of reduced accuracy, and will not be discussed in this paper}. Ghost cells are used to enforce the boundary conditions. Source terms are typically integrated over a cell using the Simpson's rule.
 
The code is mostly intended to be used as an Initial Value Problem (IVP) solver. Starting from an arbitrary initial solution, the code evolves the system of equations in time, and converges towards a steady-state, if such a solution exists. Convergence is determined by the user based on the time evolution of various macroscopic quantities, as well as by the norm of the equations' residuals. The code can also directly search for a steady-state by setting the time-derivatives to 0, through a non-linear Newton solver. This is typically run only after a temporal evolution of the equations, to provide the solver a reasonable first guess for the solution. 

 Several time-stepping schemes are implemented. SPLEND1D can either rely on the time-stepping algorithms implemented in the TS environment of the PETSc library \cite{petsc-user-ref,petsc-efficient}, using by default the fully implicit Crank-Nicolson method. In schemes that requires the computation of the Jacobian of the system, such as the Crank-Nicolson method, this operation is performed by PETSc using finite-differences with coloring. The Jacobian is typically recomputed every 10-40 iterations, a number that is set at run-time by the user. The time-step is free to evolve between user-prescribed minimum and maximum values, with PETSc taking care of the time-step adaptivity. Alternatively to the use of PETSc, and not demonstrated in this paper, SPLEND1D is equipped with a ``linearized'' IMEX (IMplicit-EXplicit) scheme, where the advection and source terms are treated explicitly, and the diffusion and viscosity operators are handled implicitly, using the nonlinear transport coefficients of the previous time-step. The implicit part can then be rewritten as a succession of tridiagonal matrix inversions, performed using the Thomas algorithm, which scales linearly with the number of cells. While this scheme comes with a Courant-Friedrichs-Lewy (CFL) condition that restricts the time-step, its linear scaling with the resolution and number of equations makes it an interesting solver for cases with a large number of cells (provided the cells are not so small that the CFL condition becomes very strict). Due to the relatively small size of the typical problem solved by SPLEND1D (around $\sim$500-1000 cells for a maximum of 8 equations, that is, $\sim$4000-8000 degrees of freedom), the code is currently not parallelized, although extension to an MPI-OpenMP parallelized code would be relatively straightforward, leveraging the capabilities of PETSc for the MPI parallelization. In section \ref{sec:basecase}, after presenting a base case used to demonstrate various features of the code, we will briefly present the performance of SPLEND1D in terms of convergence and speed. 

\section{Presentation of Base case}\label{sec:basecase}
\subsection{Parameters}
This section presents the base case investigated in this paper to illustrate in more detail the capabilities of the code, to highlight the role of various modelling parameters and to demonstrate the code's capabilities to unravel the physics behind plasma detachment. The simplified model given by equations (\ref{eq:sf_continuity})-(\ref{eq:sf_energy}) is used, assuming $T_i=T_e$ as opposed to the two-fluid model. The diffusive neutral model is employed (equations (\ref{eq:nn_diff}) and (\ref{eq:dn})), and we impose $T_n=T_i$. Symmetric boundary conditions are applied, and the velocity is imposed greater or equal than the sound speed at the right boundary. Table \ref{tab:basecase_parameters} summarizes the values of the various parameters in these simulations. $R$, $\gamma_i$, $\gamma_e$ and $\tau_n$ are based on reasonable values, whereas $L_\|$, $c_c$ and $\alpha$ are chosen based on typical TCV \cite{Reimerdes_NF2022} values. 500 grid cells are simulated, with the grid accumulation chosen so that the width of the last cell is approximately 0.8 mm in the parallel direction. These choices will be further discussed in section \ref{sec:numerics}.

\begin{table}[ht]
\begin{center}
\begin{tabular}{| c | c | c |}
\hline
Parameter & Definition & Value \\
\hline
$\ln\Lambda$ & Coulomb logarithm & Equation (\ref{eq:coulomb_logarithm_smooth}) \\
$\bar{\tau}$ & $T_i = \bar{\tau} T_e$ & 1 \\
$\gamma_i$ & Ion sheath transmission coefficient & 3.5 \\
$\gamma_e$ & Electron sheath transmission coefficient & 5.5 \\
$\alpha_{i}$ & Ion heat flux limiter & 0.6 \\
$\alpha_{e}$ & Electron heat flux limiter & 0.6 \\
$c_c$ & Carbon concentration &  $2\%$  \\
\hline
$R$ & Recycling rate & $99\%$ \\
$\tau_n$ & Neutral confinement time &  $0.05~\mathrm{ms}$  \\
\hline
$B$ & Magnetic field &  Constant  \\
$\alpha$ & Field-line angle &  $4^o$  \\
$L_\|$ & Parallel connection length & $25~\mathrm{m}$ \\
\hline
$\nu_{num}$ & Numerical viscosity &  0  \\
\hline
\end{tabular}
\caption{{\label{tab:basecase_parameters}\raggedright Summary of the parameters used for the base case simulations.}}
\end{center}
\end{table}
Since the system is source-driven, volumetric particle and energy source terms are needed. The energy sources, both for ions and electrons, are defined as Gaussian sources peaked at $s=0$ with a standard deviation (characteristic width) $\sigma=1.76~\mathrm{m}$ and amplitudes $\bar{H}_i = \bar{H}_e = 2.80~\mathrm{MWm^{-3}}$, such that 
\begin{equation}
H_e(s) = H_i(s) = \bar{H}_i \mathrm{exp}{\left( \frac{-s^2}{2\sigma^2}\right)}. 
\end{equation}
The neutral particles' source is defined as a constant source along the flux tube, such that
\begin{equation}
H_n(s) = \bar{H}_n = 1.58\times 10^{21}~\mathrm{m^{-3}s^{-1}}.
\end{equation}
The charged particles' source, $H_P$, is set to $0$.

\begin{figure}
\centering
\includegraphics[width=1.00\linewidth]{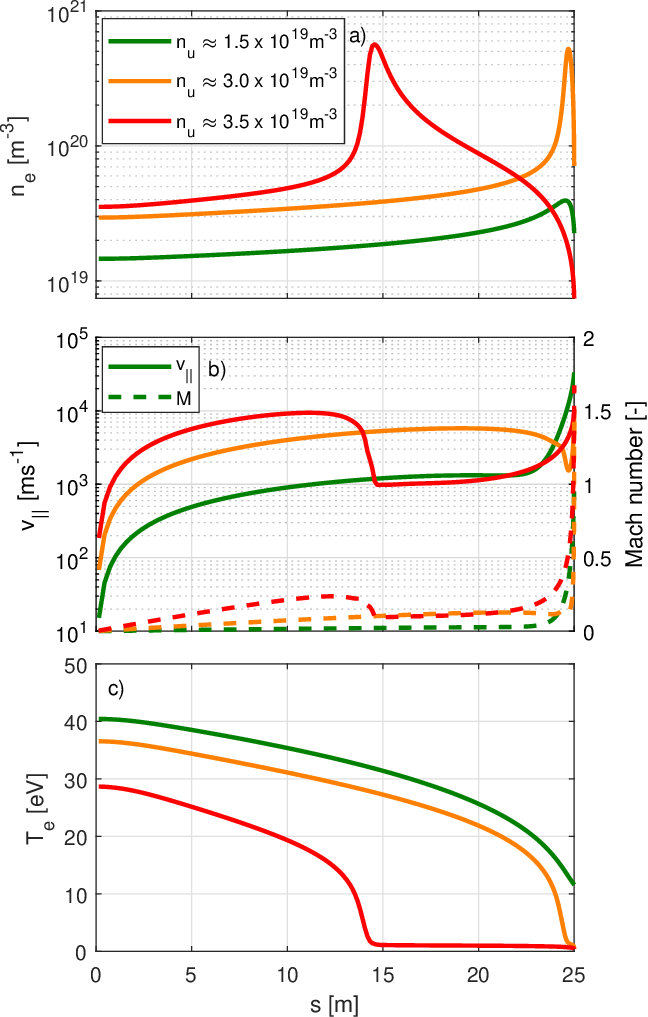}
\caption{\label{fig:n_v_T}\raggedright (a) Plasma density, (b) velocity and Mach number, and (c) temperature profiles along the SOL length, $s$, for the Base Case - in attached conditions (green), for a detached case (orange), and for a strongly detached case (red). At the target, the Mach number $M$ is $M=1$ for the attached case, $M\approx1.7$ for the detached and strongly detached cases.}
\end{figure}

\begin{figure}
\centering
\includegraphics[width=1.00\linewidth]{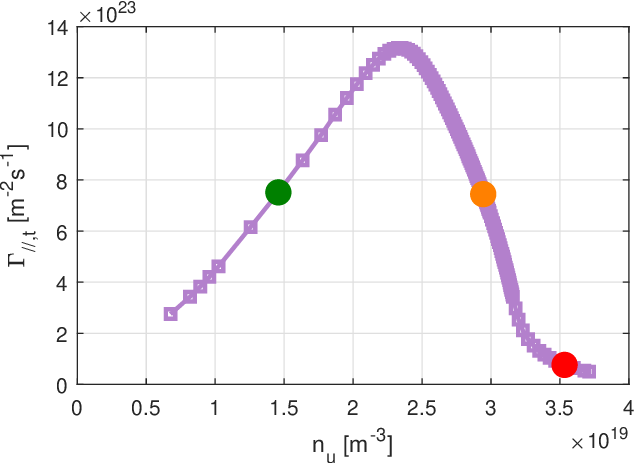}
\caption{\label{fig:Basecase_jsatrollover}\raggedright Target ion saturation current as a function of upstream density $n_u$, with the attached, detached , and strongly detached base cases of Figure \ref{fig:n_v_T} shown by the green,  orange, and red dots respectively.}
\end{figure}

This results in an upstream density of $1.5 \times 10^{19}$ m$^{-3}$ and upstream temperature of $\approx40$ eV. The full temperature and density profiles are shown in figure \ref{fig:n_v_T} in green, alongside the velocity and Mach number profiles. A target temperature of $11.6$ eV suggests that the base case represents an attached plasma regime. In this scenario, increasing the upstream plasma density via an increase in the particle source $H_n$ leads to a rollover in the target ion flux $\Gamma_{\|,t}$, Figure \ref{fig:Basecase_jsatrollover}, indicating the onset of detachment. Example profiles for detached and strongly detached cases are shown in orange and red in Figure \ref{fig:n_v_T}.

We note that in Figure \ref{fig:Basecase_jsatrollover}, each point is the steady-state result of simulations with different values of $H_n$. They were obtained by simulating $1~\mathrm{s}$ of plasma dynamics, and then applying the steady-state solver discussed in section \ref{sec:num_details}. This will be the case in all simulation results presented in this paper, except the time-dependent simulation presented in section \ref{sec:timedep}.

\subsection{SPLEND1D performance in the base case}\label{sec:numerics}
\begin{figure}
\centering
\includegraphics[width=1.00\linewidth]{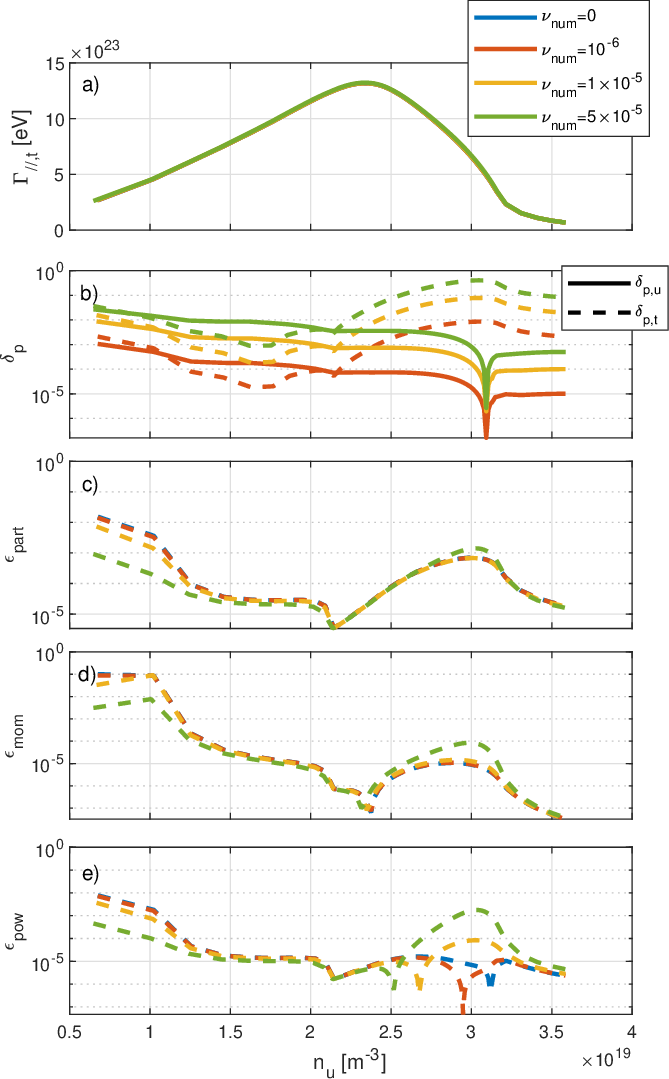}
\caption{\label{fig:Accuracy_vs_vnum} \raggedright (a) Target particle flux for different values of $\nu_{num}$, as a function of the upstream density $n_u$. (b) Effect of $\nu_{num}$ on the upstream ($\delta_{p,u}$) and target ($\delta_{p,t}$) $\delta_p$ indicator for different values of $\nu_{num}$, using $\nu_{num}=0$ as a reference, as a function of the upstream density $n_u$. (c)-(d)-(e) Effect of $\nu_{num}$ on the particle balance ($\epsilon_{part}$), momentum balance ($\epsilon_{mom}$), and energy balance ($\epsilon_{pow}$), as a function of the upstream density $n_u$.}
\end{figure}

In this section, we aim to quantify the performance of SPLEND1D in terms of accuracy in the base-case scenario described previously. In particular, we will use the methodology used in Ref.\cite{Derks_PPCF2022}, although one should note that the results are not directly comparable as the considered cases differ. Integrating Equation (\ref{eq:sf_continuity}) from upstream ($s=0$, subscript $u$) to the target ($s=L_\|$, subscript $t$) and neglecting the time derivative, one can define the numerical error of the particle balance, $\epsilon_{part}$, as
\begin{equation}
\epsilon_{part} = \left|1 - \frac{\left[\frac{nu_\|}{B}\right]^t_u}{\int^t_u\left(\frac{S_p^n + H_P}{B}\right)\mathrm{d}s} \right|.
\end{equation}
Since this paper focuses mainly on steady-state simulations, we will focus on the code performance and accuracy in such conditions. In steady-state conditions, one should ideally find $\epsilon_{part} = 0$. Any finite value of $\epsilon_{part}$ comes either from numerical errors or the numerical tolerance, as the steady-state solver has removed the time-derivatives from the system. Similarly, one can define the numerical error of the momentum balance (Equation (\ref{eq:sf_momentum})), $\epsilon_{mom}$, as 
\begin{equation}
\epsilon_{mom} = \left|1 - \frac{\left[\frac{m_inu_\|^2+p}{B}\right]^t_u}{\int^t_u \frac{-\frac{p}{B}\frac{\partial B}{\partial s}+ S^u_{p,\|} - \mathbf{b}\cdot\nabla\cdot\mathbf{\Pi}}{B}\mathrm{d}s}\right|,
\end{equation}
and the numerical error of the power balance (Equation (\ref{eq:sf_energy})), $\epsilon_{pow}$, as 
\begin{equation}
\epsilon_{pow} = \left|1 - \frac{\left[q_{\|}^{tot}\right]^t_u}{B\int^t_u \frac{S_i^E + S_e^E + S_{imp}^E + H_e + H_i}{B}\mathrm{d}s}\right|, 
\end{equation}
where 
\begin{equation}
q_{\|}^{tot} = \frac{5}{2}n\left(1+\bar{\tau}\right)T_eu_\| + \frac{1}{2}m_i n u^3_\| + q_{\|,i}^{cond} + q_{\|,e}^{cond} + q_{\|,i}^{visc}.
\end{equation}
We also define $\delta_p$, as
\begin{equation}
\delta_p = \left| \frac{p - p_{\mathrm{ref}}}{p_{\mathrm{ref}}} \right|,
\end{equation}
where $p = 2nT$ is the static pressure and $p_{\mathrm{ref}}$ the static pressure of a reference simulation. $\delta_{p}$ is defined either upstream ($s=0$) or at the target ($s=L_\|$). $\delta{p}$ quantifies the variation of the static pressure in each simulation to that of a reference simulation. This will be used later in this section to assess the effect of target-cell width and number of cells on the numerical convergence of SPLEND1D. We now evaluate these quantities for the density ramp shown in figure \ref{fig:Basecase_jsatrollover}. Figure \ref{fig:Accuracy_vs_vnum} shows the numerical error on the particle, momentum and energy balances, $\epsilon_{part}$, $\epsilon_{mom}$, $\epsilon_{pow}$. Two regimes can be identified. For $n_u>1.5\times10^{19}~\mathrm{m^{-3}}$, SPLEND1D shows excellent convergence properties even in the absence of numerical viscosity $\nu_{num}$ (see equation (\ref{eq:par_visc})) , with a maximum error on the particle balance of $\epsilon_{part}^{max}\approx 0.1\%$, while it is $\epsilon_{mom}^{max}\approx 0.0015\%$ for the momentum and $\epsilon_{pow}^{max}\approx 0.0023\%$ for the power balance. For $n_u<1.5\times10^{19}~\mathrm{m^{-3}}$, the situation is more intricate, and while particle and power balances remain satisfactory, the error on the momentum balance can become important (up to $\approx30\%$) in the absence of numerical viscosity $\nu_{num}$. This is due to the very strong velocity gradient that will form just in front of the target to bring the flow from a very low value to the sound speed. This can be alleviated by the addition of a finite $\nu_{num}$, bringing the momentum error below $1\%$, Figure \ref{fig:Accuracy_vs_vnum}d. This has little impact on the overall outputs of the simulation, Figure \ref{fig:Accuracy_vs_vnum}a, where $\Gamma_{\|,t}$ remains virtually unchanged across the various values of $\nu_{num}$. Similarly, the upstream and target pressure are only affected by up to $3\%$ by the addition of $\nu_{num}$, Figure \ref{fig:Accuracy_vs_vnum}b. Another possibility would be to increase the resolution of the grid near the target, for instance by reducing the width of the cells, as will be shown later in this section. Overall, these results demonstrate that the numerical accuracy of SPLEND1D across the different regimes is satisfactory, from attached to detached regimes. 

\begin{figure}
\centering
\includegraphics[width=1.00\linewidth]{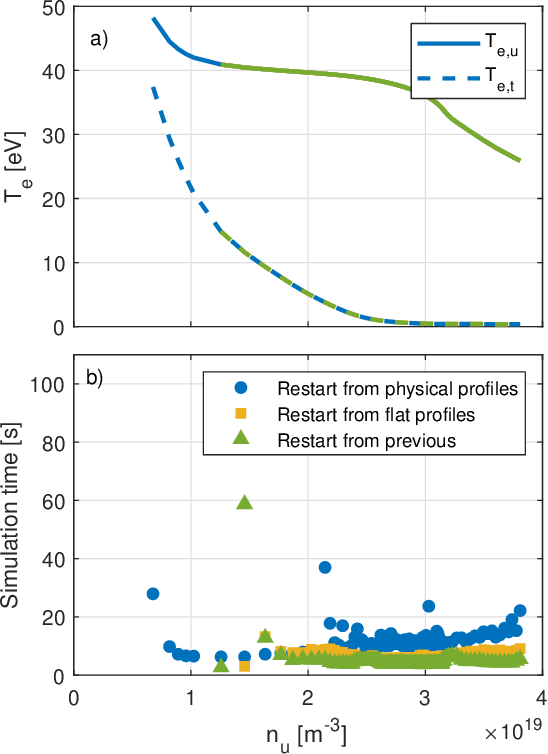}
\caption{\label{fig:density_ramps_vs_strategy_and_error}\raggedright Results of simulations using the three strategies outlined in section \ref{sec:numerics}. (a) Upstream (solid) and target (dashed) temperature as a function of upstream density $n_u$, with indistinguishable results across the applied strategies. (b) Simulation run-time required for $1~\mathrm{s}$ of simulated plasma time.}
\end{figure}

We next explore the performance, in term of computing time, of SPLEND1D, across a density ramp with the same input parameters as that of Figure \ref{fig:Basecase_jsatrollover}. For this, we evaluate the simulation time, towards steady-state, for different values of $\bar{H}_n$, applying three different strategies:
\begin{enumerate}
\item The simulations are run sequentially, each starting from the steady-state obtained from the previous simulations. 
\item Each simulation is run independently, starting from ``physical" profiles obtained from a steady-state simulation with $\bar{H}_n = 1.58\times 10^{21}~\mathrm{m^{-3}s^{-1}}$.
\item Each simulation is run independently, starting from flat, unphysical profiles ($n=2.0\times10^{19}~\mathrm{m^{-3}}$, $T_e = T_i = 20~\mathrm{eV}$, $u_\| = 0~\mathrm{ms^{-1}}$). 
\end{enumerate}
In all simulations presented in this section, the same convergence parameters for the PETSc solver are used. The time-step $\Delta t$ is allowed to vary between $\Delta t = 5.71\times 10^{-11}~\mathrm{s}$ and $\Delta t = 5.71\times 10^{-4}~\mathrm{s}$. It is automatically adapted by PETSc based on convergence and error estimates. The simulations are run for $1~\mathrm{s}$ of simulated plasma time, after which the steady-state solver is applied to find an exact (within nonlinear solver tolerance) steady-state solution. All simulations were performed on a typical laptop, with an Intel\textregistered Core\texttrademark i7-8565U CPU. These three simulation strategies provide identical output profiles (within the nonlinear solver tolerance), Figure \ref{fig:density_ramps_vs_strategy_and_error}a. 
Figure \ref{fig:density_ramps_vs_strategy_and_error}b reports the simulation run-times. Starting from flat, unphysical profiles (strategy 3), all simulations converge within $\approx~$20~s, except for a few outliers that require up to 40s. Starting from physical profiles of an attached case (strategy 2) yields even lower computation times, approximately 10s, except at low densities (lower than the initial simulation), where simulations can take relatively long times or even fail to converge in less than 10min (after which the simulations were stopped). A similar observation is done for the sequential scan of $\bar{H}_n$ (strategy 1), restarting from the previous simulation (with a slightly different value of $\bar{H}_n$), where the simulation time can drop to $\approx~$6~s, except at low density where simulations can struggle to converge in reasonable time. In conclusion, this section demonstrates the numerical performances of SPLEND1D, which can achieve convergence towards a steady-state in less than 30s across a wide range of regimes, including the strongly detached one. 

We now inspect the dependence of SPLEND1D results and accuracy on the grid resolution, for simplicity in the case of a symmetric domain, as done in the base case. The resolution of the grid is controlled by two parameters, the total number of cells and the width of the last cell before the target, thereafter denoted $h_{end}$, which define entirely the grid, equation (\ref{eq:cellsize}). In the following, we will assess the effect of both these parameters in the attached base case, varying only one parameter at a time. For simplicity, we group the error on particle balance, momentum balance and power balance under a single new term, $\epsilon_{tot}$, defined as 
\begin{equation}
    \epsilon_{tot} = \sqrt{ \epsilon_{part}^2 + \epsilon_{mom}^2 +\epsilon_{pow}^2}.
    \label{eq:epsilon_tot}
\end{equation}

\begin{figure}
\centering
\includegraphics[width=1.00\linewidth]{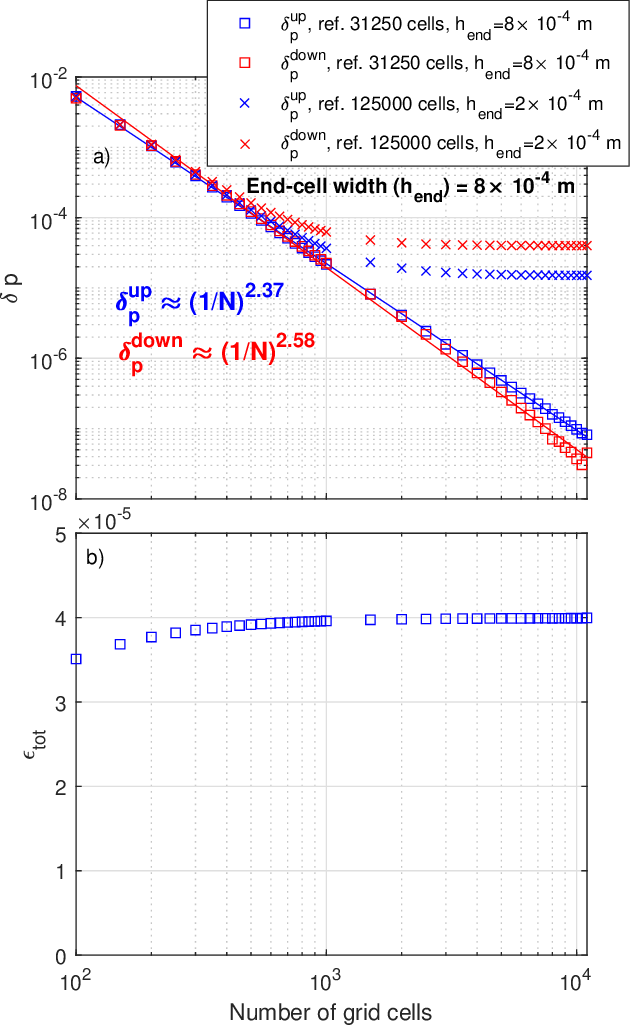}
\caption{\label{fig:error_vs_numcells}\raggedright (top) $\delta_{p}$ for the upstream pressure (blue) and downstream pressure (red), using either the 31250 cells case as reference (squares) or the 125000 cells as references (crosses). The scaling relations of $\delta_p$ with $N$ are computed using the 31250 cells case as reference; (bottom) Error of the particles, momentum, and energy balances $\epsilon_{tot}$ as a function of the number of grid cells, for a fixed width of the last cell, $h_{end}=0.8~\mathrm{mm}$.}
\end{figure}

We start by considering the influence of the number of grid cells, $N$, keeping $h_{end}=0.8~\mathrm{mm}$. $N$ is varied from 100 to 1000 by increments of 50, and then from 1000 to 10000 by increments of 500, Figure \ref{fig:error_vs_numcells}. All simulations converged. We first use a simulation with 31250 cells of 0.8~mm each as a reference for the evaluation of $p_{ref}$. As $N$ is increased, we observe that both the upstream and downstream values of $\delta_{p}$ decrease quadratically with $N$, as one could expect from a second order scheme. $\epsilon_{tot} $ remains largely unaffected. Further, using now a high resolution (125000 cells, $0.2~\mathrm{mm}$ each) simulation as reference, we find that $\delta_{p}$ stagnates when the number of cells is higher than 2000. This is because, as will be shown in the next paragraph, the accuracy of SPLEND1D is largely dictated by the width of the cells close to the target, kept fixed in this scan. Hence, increasing further the number of cells does not lead to an increase of the accuracy. Furthermore, and interestingly, even at fairly low resolution ($\sim100$ cells), $\delta_p$ for the upstream and target pressure remain lower than $1\%$. 

\begin{figure}
\centering
\includegraphics[width=1.00\linewidth]{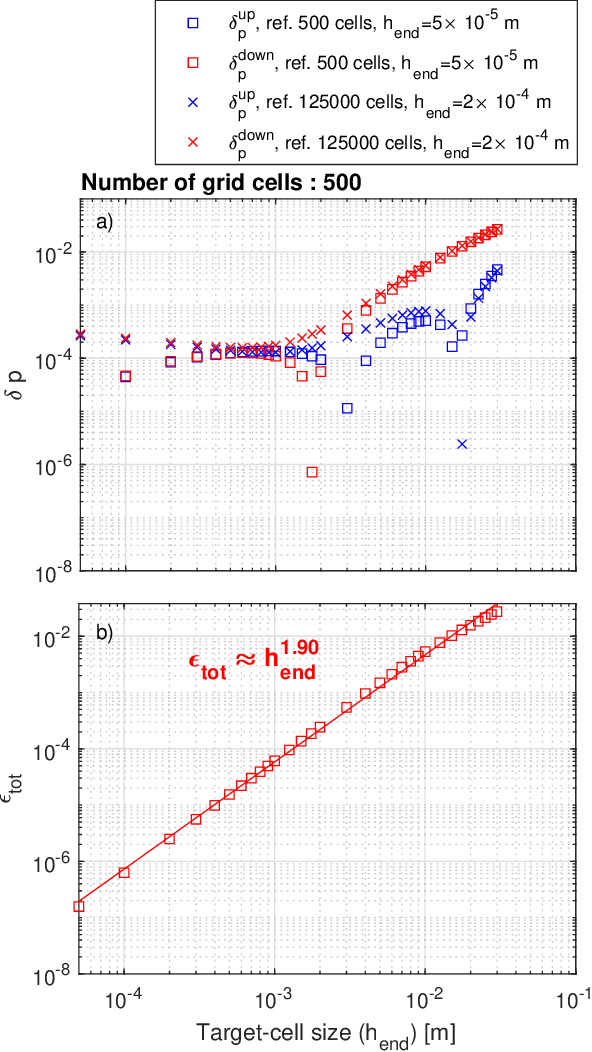}
\caption{\label{fig:error_vs_cellsize}\raggedright (top) $\delta_{p}$ for the upstream pressure (blue) and downstream pressure (red) , using either the 500 cells case as reference (squares) or the 125000 cells as references (crosses); (bottom) Error of the particles, momentum and energy balances $\epsilon_{tot}$ as a function of the width of the last cell, $h_{end}$, for a fixed number of grid cells.}
\end{figure}

We now keep the number of grid cells constant (500 cells) whilst changing $h_{end}$. As $h_{end}$ is reduced, $\epsilon_{tot}$ and $\delta_p$ (both upstream and downstream) strongly decrease, Figure \ref{fig:error_vs_cellsize}. This indicates that the accuracy of SPLEND1D largely depends on the width of cells near the target. In particular, $\epsilon_{tot}$ scales approximately as $h_{end}^2$.

Taken together, the results of Figures \ref{fig:error_vs_numcells} and \ref{fig:error_vs_cellsize} show that the accuracy of SPLEND1D is largely dictated by the grid resolution near the targets, with the total number of grid cells playing a lesser role. However, as the number of grid cells is increased, or the width of the last cell is decreased, the computational cost increases. From Figures \ref{fig:density_ramps_vs_strategy_and_error}, \ref{fig:error_vs_numcells}, and \ref{fig:error_vs_cellsize}, the resolution chosen for the base case (500 grid cells, 0.8mm target cell width) appears to be a good trade-off between accuracy and numerical cost, and is retained for the rest of the simulations presented in this paper. 

\subsection{The two-point model formatting}
To evaluate the importance of physical momentum and power loss processes in the base case simulations, we first introduce briefly the two-point model formulation. It is a model that does not consider the spatial distribution of plasma parameters along the SOL, but considers quantities at only two points: upstream and target, where upstream can be any point along the flux tube. Furthermore, it considers a steady state situation. 

Momentum losses between upstream and target are grouped into a single momentum loss factor, defined as the relative difference in upstream and target total pressures. {We remark here  that some works\cite{stangeby2000plasma} define the momentum loss factor as $(1-f_{mom})$, being simply the ratio of target to upstream pressure.},
\begin{align} \label{eq:fmom}
    f_{\text{mom}}&= \frac{p_{tot,u} - p_{tot,t}}{p_{tot,u} }\\&=1-\frac{n_tT_{e,t}}{n_uT_{e,u}}\frac{(1+ \frac{T_{i,t}}{T_{e,t}})}{(1+ \frac{T_{i,u}}{T_{e,u}})}\frac{(1+M_t^2)}{(1+M_u^2)},
\end{align}
where $M_t$, $M_u$ are the target and upstream Mach numbers respectively: $M=v_{\|}/c_s$. Similarly, volumetric power losses along the SOL are described by a single power loss factor, $f_{\text{power}}$. However, because the spatial location of the power sources in SPLEND1D can be chosen arbitrarily, care must be taken for the definition of $f_{\text{power}}$. We start by integrating, from the target to upstream, the sum of the electron and ion energy equations, equations (\ref{eq:eenergy}) and (\ref{eq:ienergy}), taken at steady-state. Defining the heat flux as the sum of the convective, conductive, and viscous contributions, 
\begin{equation}
    q_{\|}^{tot}=\frac{5}{2}n(1+T_i/T_e)T_e u_{\|}+\frac{1}{2}m_inu_{\|}^3+q^{cond}_{\|,i}+ q^{cond}_{\|,e} + q_{\|,i}^{visc},
\end{equation}
we then have 
\begin{equation}
    B\frac{\partial}{\partial s} \left(\frac{q_{\|}^{tot}}{B}\right) \nonumber \\ = S_i^E + S_e^E + S_{imp}^E + H_e + H_i.
    \label{eq:sum_energ_equations}
\end{equation}
We now define the total energy loss factor, $f_{pow}$, as 
\begin{equation}\label{eq:fpower}
    f_{pow} = 1-\frac{q_{\|,t}^{tot}B_u/B_t}{q_{in}}, 
\end{equation}
where 
\begin{equation}\label{eq:qin}
    q_{in}=q_{\|,u}^{tot}+B_u\int^t_u\frac{(H_e+H_i)}{B}\mathrm{d}s, 
\end{equation} and  
\begin{align}
q_{\|,t}^{tot}&=q^{cond,e}_{||,t}+q^{cond,i}_{||,t}+q^{conv,t}_{||,u}+q^{conv,t}_{||,u}+q_{\|,i}^{visc,t}\\&=n_tu_{||}(\gamma_eT_e+\gamma_iT_i).
\label{eq:bcs_2PM}
\end{align} 
When applied to the SPLEND1D model described in section \ref{sec:plasma_eq}, the contribution of each source term to the momentum and power losses can be evaluated such that the important processes can be identified. The breakdown of the momentum and power loss factors is given explicitly in appendix \ref{app:2PM_fmom_fpower}. Figure \ref{fig:IonizationVsRecombination} shows the individual contributions to $f_{pow}$ and $f_{mom}$, including atomic sources and viscosity. This will be studied in further detail in section \ref{sec:rollover_processes}. {We remark here that, with this definition of $f_{mom}$, geometry effects related to total flux expansion are embedded within volumetric source and sink terms. In order to highlight more clearly the role of total flux expansion on target conditions, it can be preferable to use the $f_{mom}$ definition proposed in Ref. \cite{Carpita_Arxiv2023}, which is specifically formulated to elucidate the total flux expansion effect. We further remark that, since the viscous heat flux is included in the expression of $q_{\|}^{tot}$ and enters the heat flux boundary condition (equation \ref{eq:bcs_2PM}), the contribution of viscosity is not included in $f_{pow}$.}

Considering particle, momentum and power balances with the various loss factors, the target temperature and density can be expressed as a function of upstream total pressure, $p_{tot,u}$, and input heat flux $q_{in}$ (as defined in equation (\ref{eq:qin})) as follows,
\begin{align}
    T_{e,t}&=m_i\frac{(1+M_t^2)^2}{M_t^2}\frac{(1-f_{\text{power}})^2}{(1-f_{\text{mom}})^2}\left( \frac{B_t}{B_u} \right)^2 \\ 
    & \times \frac{\left( 1+ \frac{T_{i,t}}{T_{e,t}} \right)}{\left( \gamma_e+\gamma_i\frac{T_{i,t}}{T_{e,t}} \right)^2} \frac{q_{in}^2}{p_{tot,u}^2},
\end{align}
\begin{align}
    n_t&=\frac{1}{m_i}\frac{M_t^2}{(1+M_t^2)^3}\frac{(1-f_{\text{mom}})^3}{(1-f_{\text{power}})^2}\left( \frac{B_u}{B_t}\right)^2 \\
    & \times \frac{\left( \gamma_e+\gamma_i\frac{T_{i,t}}{T_{e,t}} \right)^2}{\left( 1+ \frac{T_{i,t}}{T_{e,t}} \right)^2}\frac{p_{tot,u}^3}{q_{in}^2}.
\end{align}
This form of the two-point model, labelled as 2-point formatting \cite{Moulton_PPCF2017,Stangeby_PPCF2018}, is a reformulation of the SPLEND1D equations for the target temperature and density, given the power loss and momentum loss factors defined in equations (\ref{eq:fpower}) and (\ref{eq:fmom}).

\section{Detachment onset via density ramp in the base case}\label{sec:detachment_basecase}

\subsection{Observation of a target ion flux roll-over and onset of a total pressure drop}
\begin{figure}
\centering
\includegraphics[width=1.00\linewidth]{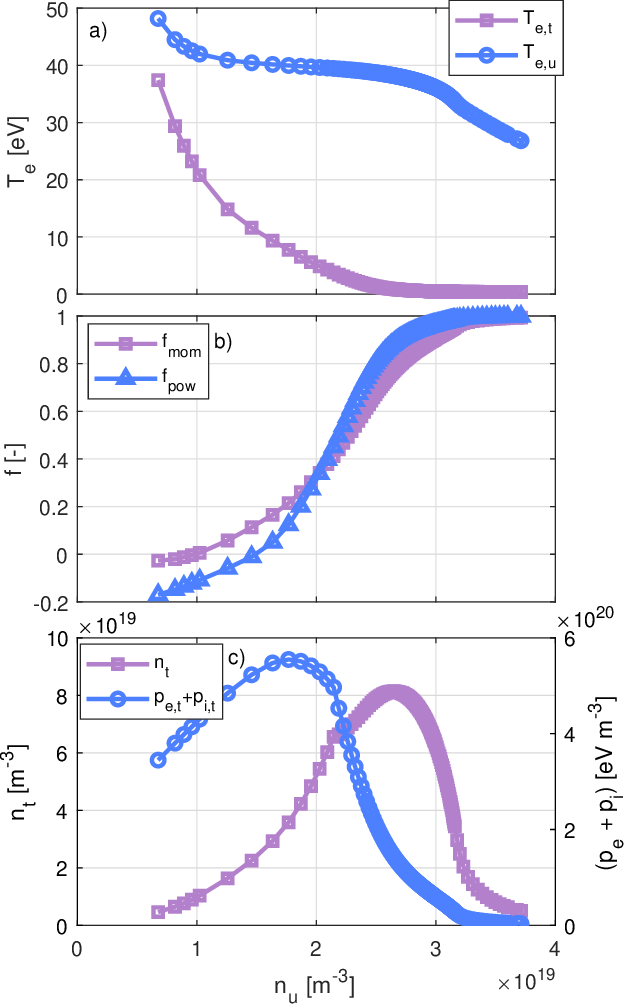}
\caption{\label{fig:Basecase_rollover}\raggedright (a) Evolution of target temperature $T_{e,t}$ and upstream temperature $T_{e,u}$ as a function of $n_u$
(b) Momentum loss factor $f_{mom}$ and power loss factor $f_{pow}$ between upstream and target, as defined in equations (\ref{eq:fmom}) and (\ref{eq:fpower}). (c) Evolution of target pressure (electrons+ions) and target density $n_t$ as a function of $n_u$.}
\end{figure}

As mentioned earlier, in the base case presented in section \ref{sec:basecase}, increasing the upstream plasma density via an increase in neutral particle source $H_n$ leads to a rollover in the target ion flux, Figure \ref{fig:Basecase_jsatrollover}, indicating the onset of detachment. The rollover is accompanied by a reduction of target temperature to less than $1~\mathrm{eV}$ (Figure \ref{fig:Basecase_rollover}a). In contrast, the upstream temperature $T_u$ is much less sensitive, and only at the highest degree of detachment does it start to degrade significantly, dropping from approximately $\approx40$ eV to $\approx28$ eV. The particle flux rollover is accompanied by a pressure drop, as shown by evaluating $f_{mom}$ (equation (\ref{eq:fmom})), which increases from 0 (no pressure drop) at low density to 0.99 at the highest density achieved in the present simulations, Figure \ref{fig:Basecase_rollover}b. The plasma power loss factor, $f_{pow}$ (equation (\ref{eq:fpower})), exhibits a similar behaviour as the momentum loss with increasing density, but reaches saturation ($f_{pow}\sim 1$) at a lower upstream density. This suggests that the onset of detachment is first driven by the increase of power losses, that precedes momentum losses. 

\subsection{Investigate process at play} \label{sec:rollover_processes}
\begin{figure*}
\centering
\includegraphics[width=1.00\linewidth]{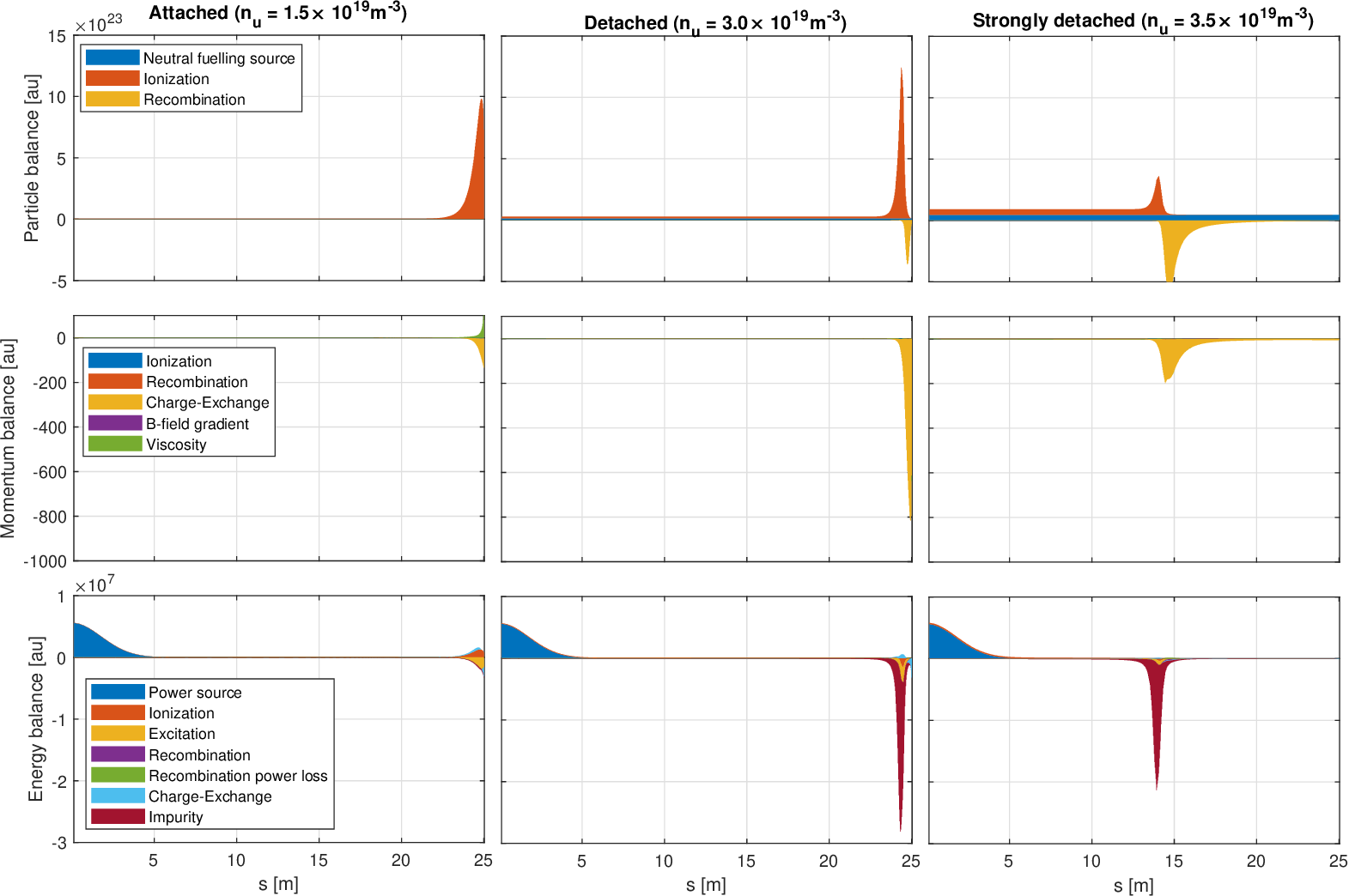}
    \caption{\raggedright Source ans sink terms contribution to plasma (top) particle, (middle) momentum and (bottom) energy equations (equations (\ref{eq:sf_continuity}), (\ref{eq:sf_momentum}) and (\ref{eq:sf_energy}) respectively), in (left) an attached case, (middle) a detached case, (right) a strongly detached case. In the energ balance, the "recombination" term corresponds to the loss of the ion energy while the "recombination power loss" corresponds to the loss of electron energy.}
    \label{fig:partmomenergy_breakdown}
\end{figure*}

\begin{figure}
\centering
\includegraphics[width=1.00\linewidth]{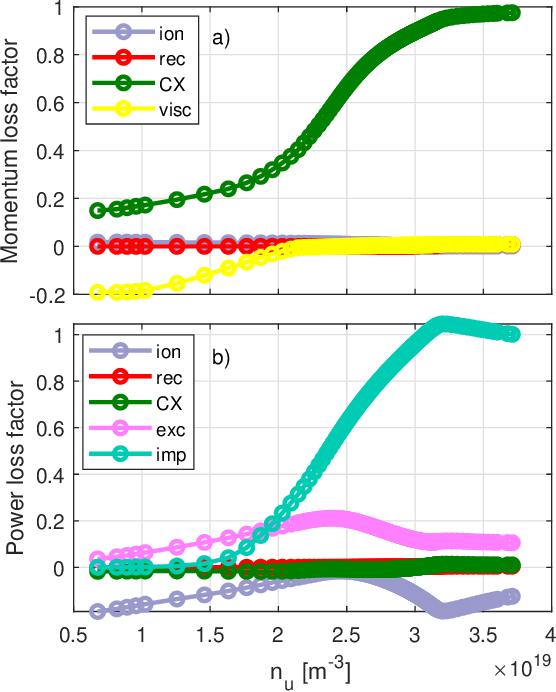}
\caption{\label{fig:IonizationVsRecombination}\raggedright Source terms contributing to plasma (top) momentum loss and (bottom) power loss factors formulated as in equations (\ref{eq:fmom}) and (\ref{eq:fpower}), and appendix \ref{app:2PM_fmom_fpower}, as a function of upstream electron density $n_u$. The momentum loss factor includes losses due to ionization (ion), recombination (rec), charge exchange (CX) and viscosity (visc). The power loss factor includes losses due to ionization (ion), recombination (rec), charge exchange (CX), excitation (exc) and impurity radiation (imp). Negative values represent a gain in the plasma momentum or energy.}
\end{figure}

The first sign of detachment, Figure \ref{fig:Basecase_rollover}, appears to be a target electron pressure rollover, along with a target ion flux rollover and a target temperature decrease to below $1~$eV. With a further increase in upstream density, the target electron density rolls over, followed by a saturation in the power losses and then momentum losses. The underlying processes behind these features can be studied with SPLEND1D, which directly outputs each term contributing to the particle, momentum and energy balance equations (equations (\ref{eq:sf_continuity}), (\ref{eq:sf_momentum}) and (\ref{eq:sf_energy})), such that the importance of each process can be compared. Figure \ref{fig:partmomenergy_breakdown} shows the profile of each of these terms along the flux tube for an attached, a detached, and a strongly detached case. This allows to compute the contribution of each of these terms to the momentum and power losses, Figure \ref{fig:IonizationVsRecombination}. We must note that the choice of neutral model will affect the role of atomic processes in momentum and power losses. The simulations discussed here employ the diffusive neutral model with $T_n=T_i$, and so the power loss due to charge exchange may be underestimated. This is discussed in further detail in section \ref{sec:freeparam_neutralmodel}.

The target electron pressure rolls over as the momentum losses, which are dominated by charge exchange reactions, begin to increase significantly, resulting from a strong increase in electron density in front of the target, as well as from a drop in temperature, which favors charge-exchange reactions over ionization. This leads to the rollover of the target ion flux. The region in front of the target becomes much cooler, and a region of strong temperature gradient moves upstream, with $T_{e,t}<1~$eV. This front movement is also seen in other atomic processes (ionization, recombination and impurity radiation, figure \ref{fig:partmomenergy_breakdown}), and eventually in the plasma density front as strong detachment is achieved, leading to a rollover in target electron density. The impurity radiation increases as the divertor becomes cooler and denser, becoming the dominant power loss mechanism, until $f_{pow}$ saturates. As the neutral density continues to increase in the cool divertor, charge exchange momentum losses continue to increase until $f_{mom}$ also saturates at $f_{mom}\sim1$. Note that although momentum and power losses due to recombination increase with upstream density, they remain negligible compared to other atomic sources throughout the density range explored. However, recombination appears as a significant contributor of the particle balance, Figure \ref{fig:partmomenergy_breakdown}.

\begin{figure}
\centering
\includegraphics[width=1.00\linewidth]{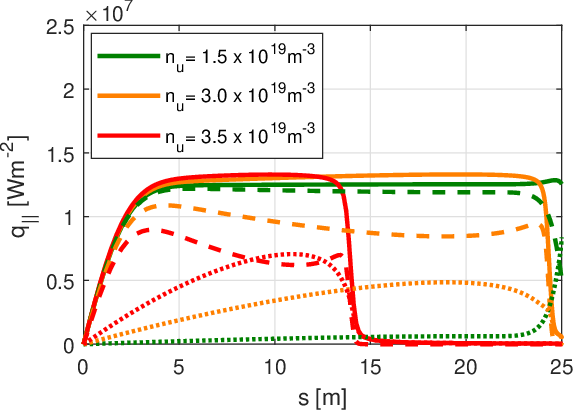}
\caption{\label{fig:heatfluxprofs}\raggedright Parallel heat flux profiles for the base case with $n_u=1.5\times10^{19} ~\mathrm{m}^{-3}$ (green), for a detached case with $n_u=3.0\times10^{19} ~\mathrm{m}^{-3}$ (orange), and for a strongly detached case with $n_u=3.5\times10^{19} ~\mathrm{m}^{-3}$. The convective (dotted) and conductive (dashed) contributions are plotted alongside the total parallel heat flux (solid).}
\end{figure}

One of the key assumptions of the standard Two-Point Model (TPM) that is often used as a first model to study SOL physics \cite{stangeby2000plasma}, is that the heat transport is mostly due to (electron) heat conduction. Whilst convection can be enabled in the extended TPM through some ad-hoc $f_{cond}$ parameter, the TPM itself does not provide a self-consistent way to estimate the value of $f_{cond}$. In SPLEND1D, we find that convective heat transport in the current simulations is small but non-negligible, Figure \ref{fig:heatfluxprofs}. The difference between the SPLEND1D results and the TPM predictions is small at low upstream density ($n_u=1.6\times10^{19}~\mathrm{m}^{-3}$) but becomes stronger as upstream density is increased, where the convective heat flux becomes significant.

\section{Exploring the role of free parameters }\label{sec:freeparam}
As mentioned in the introduction and in the derivation of the model, 1D codes come with strong assumptions and, as illustrated by table \ref{tab:basecase_parameters}, free parameters that may affect the results of the simulations. These become especially important when attempting to use the results of such a model to explain experimental observations (although, in the case of interpretative simulations, some of them can be constrained by experimental data). It can therefore be important to understand how sensitive the code results are to these free parameters and assumptions. In this section, we review the impact of some of the main assumptions or parameters on the observations of section \ref{sec:detachment_basecase}. 

\subsection{Carbon concentration}
\begin{figure}
\centering
\includegraphics[width=1.00\linewidth]{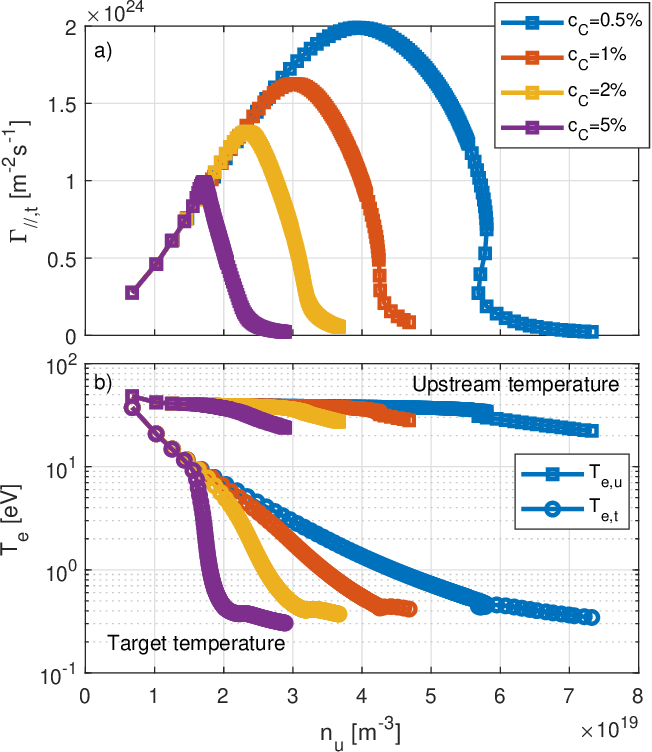}
\caption{\label{fig:SPLEND1D_Carbon_role}\raggedright (top) Target ion saturation current and (bottom) upstream and target electron temperatures as a function of $n_u$ and for different carbon concentrations.}
\end{figure}

Let us first examine the role of a parameter strongly affecting the plasma energy sink: the carbon concentration $c_c$. As expected, $c_c$ has a strong impact on the target parallel particle flux (Figure \ref{fig:SPLEND1D_Carbon_role}a), with the upstream density required for the $\Gamma_{\|,t}$ rollover decreasing strongly with $c_c$. This is due to the strong cooling induced by increased carbon radiation, leading to lower target temperatures for a given upstream density and temperature (Figure \ref{fig:SPLEND1D_Carbon_role}b), and hence a facilitated access to detachment.

\subsection{Neutral confinement time}\label{sec:freeparam_neutralmodel}
\begin{figure}
\centering
\includegraphics[width=1.00\linewidth]{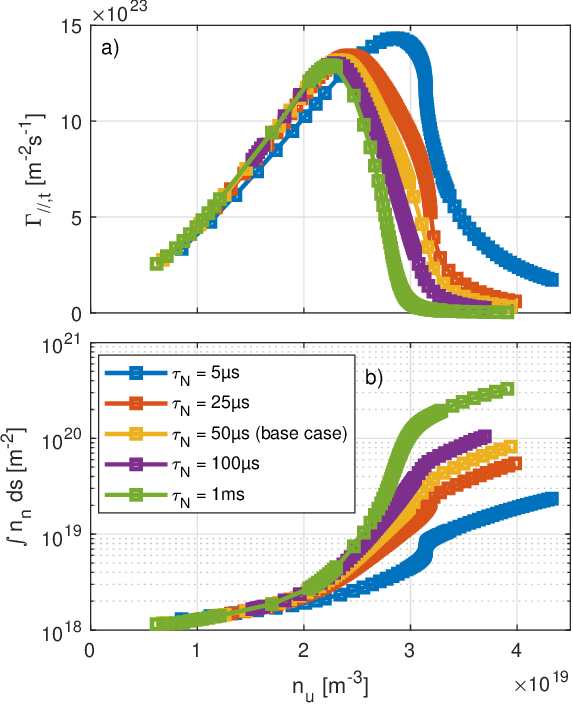}
\caption{\label{fig:SPLEND1D_TauN_role}\raggedright (top) Target ion saturation current and (bottom) integrated neutral density along the field line as a function of $n_u$, for a range of neutral confinement times $\tau_N$.}
\end{figure}

The neutral model implemented in SPLEND1D essentially features three free parameters: the neutral confinement time $\tau_N$ (see equation (\ref{eq:neutral_dens_})), the choice of neutral model and neutral temperature implementation. In this section, we focus on the free parameter $\tau_N$. Figure \ref{fig:SPLEND1D_TauN_role} plots the effect of varying $\tau_N$ on the inferred target ion flux and integrated neutral density in the base case density ramp. This reveals that, as expected, $\tau_N$ has little effect at low density, where the neutral density is low in all cases. Only at high $n_u$ does the effect of $\tau_N$ become significant: the neutral density integrated along the flux tube increases strongly with $\tau_N$, associated with a strong decrease of the target ion flux. The upstream density at which the target ion flux rolls over is not strongly affected by $\tau_N$, except for the most extreme case where $\tau_N$ has been decreased by a factor 10. 

\subsection{Heat flux limiter}\label{sec:flux_limiter}
\begin{figure}
\centering
\includegraphics[width=1.00\linewidth]{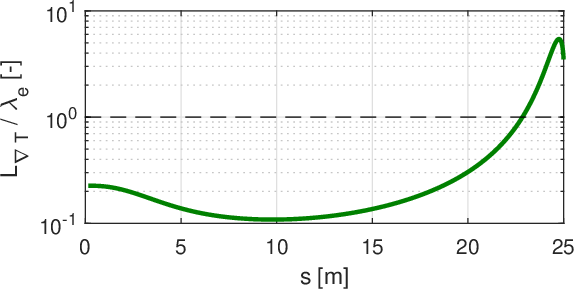}
\caption{\label{fig:collisionality}\raggedright Profile of the ratio of the parallel temperature gradient scale length to the electron collisional mean free path\cite{stangeby2000plasma} along the SOL length, $s$, for the Base Case.}
\end{figure}

The Spitzer-Harm heat flux can overestimate the physical heat flux at low plasma collisionality, when the electron mean free path, $\lambda_e$, is large compared to the electron temperature gradient scale length, $L_{\nabla T} = (\left|\nabla T_e / T_e\right|)^{-1}$. Figure \ref{fig:collisionality} shows the ratio of these scale lengths as a function of distance along the SOL, for the base case density and temperature profiles shown in figure \ref{fig:n_v_T}. At the target, the electron collisionality is high enough that the classical Spitzer-Härm heat fluxes are expected to fairly accurately predict the physical heat flux, without the need for a flux limiter. However, the upstream electron mean free path exceeds the temperature gradient scale length and so the effect of enforcing a heat flux limiter should be studied.

\begin{figure}
\centering
\includegraphics[width=1.00\linewidth]{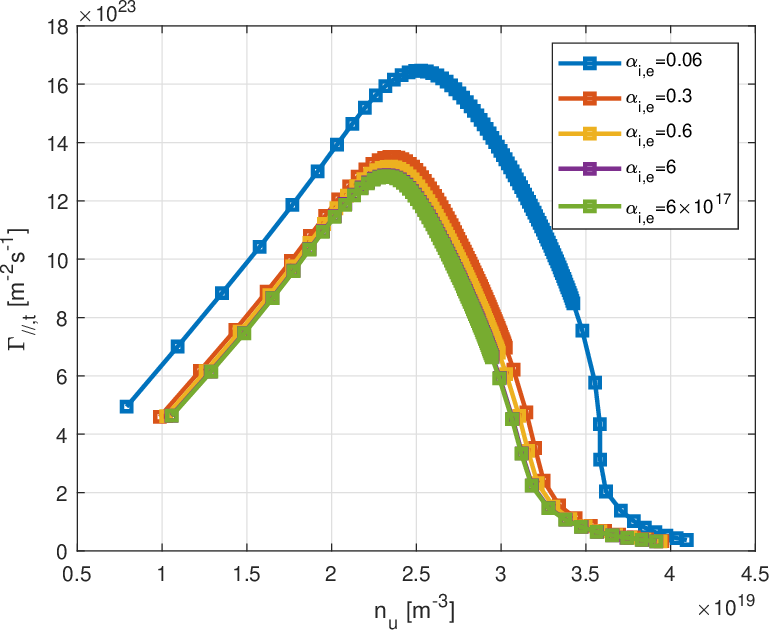}
\caption{\label{fig:Qlim_rollover}\raggedright Target ion flux rollover for the Base Case with varied heat flux limiter coefficients $\alpha_{i,e}$.}
\end{figure}

The strength of the heat flux limiter can be controlled with the parameter $\alpha_{\{i,e\}}$. Target conditions are compared here for values of $\alpha_{i}=\alpha_{e}=[0.06,0.3,0.6,6,6\times 10^{17}]$, thus testing values well below and above the typical value of 0.5 used in the literature (Ref. \cite{Day_CPP1996,Fundamenski_PPCF2005}). This includes $\alpha \gg 1$, approaching the situation without heat flux limiters. The results are found to be little affected by this choice. The target ion flux rollover and thus the detachment threshold is weakly affected by the choice of $\alpha_{\{i,e\}}$, Figure \ref{fig:Qlim_rollover}, except for the smallest value of $\alpha_{\{i,e\}}=0.06$. For $\alpha_{\{i,e\}}\geq0.3$, the magnitude of the target ion flux varies by approximately $10\%$ within the range of $\alpha_{\{i,e\}}$ studied, Figure \ref{fig:Qlim_rollover}, and the pressure drop along the SOL to the target is negligible (not shown). The target density is largely unaffected, and the target temperature is only affected at low upstream density ($\sim2\times 10^{19}\mathrm{m^{-3}}$), not shown. These results hold for a range of input power levels.

Due to the apparent insensitivity of target parameters to the heat flux limiter coefficient for the base case conditions, and for reasonable values of $\alpha_{\{i,e\}}\geq 0.3$, and over a large part of the base case density ramp, $\alpha_{\{i,e\}}$ was set to $0.6$ in all simulations presented in this paper, unless stated otherwise. %Note that the viscous flux limiter is not required either. 

\subsection{Bohm boundary condition}\label{sec:bohm_boundary_conditions}
 SPLEND1D models the plasma along a flux tube up to the entrance of the sheath ({in the case of small target angles, this actually corresponds to the entrance of the magnetic pre-sheath)}, which acts as a perfect sink that absorbs all incoming ions. This results in the so-called Bohm boundary condition, that is, for a purely parallel flow, 
\begin{equation}
    u_\| \ge c_s,
    \label{eq:bohm_condition_discussion}
\end{equation}
where $c_s$ is the sound speed. In the base case presented so far, we indeed allowed for $u_\| \geq c_s$ ($M\geq1$) at the sheath entrance. However, other codes sometimes enforce the strict equality $u_\| = c_s$, thus precluding the presence of supersonic flows at the target. In this section, we briefly discuss how this may influence the various target parameters. 

\begin{figure}
\includegraphics[width=1.00\linewidth]{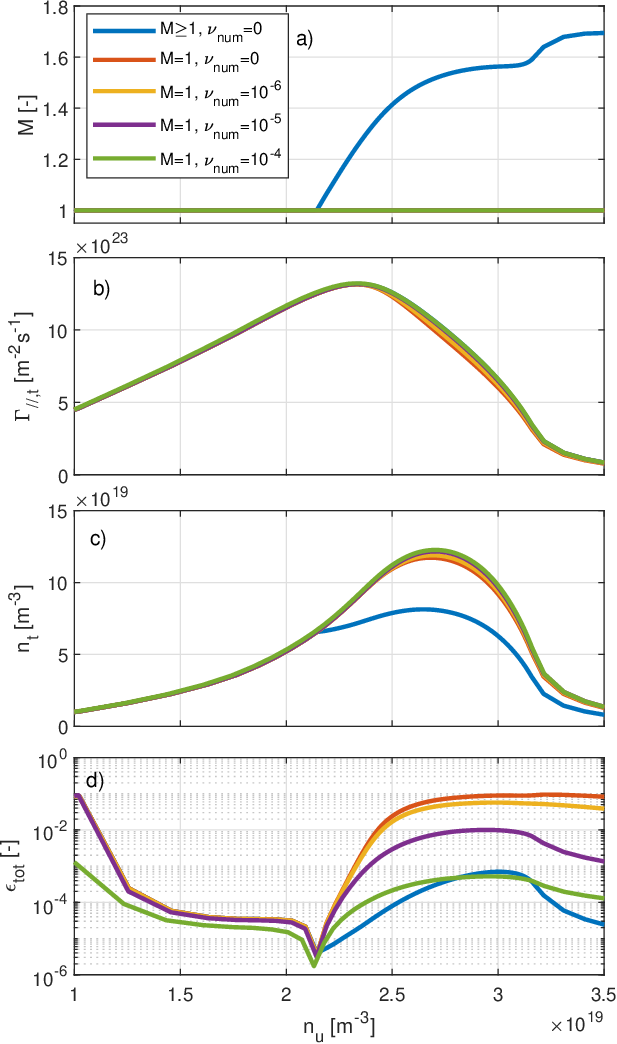}
\caption{\label{fig:cs_vs_above_s}\raggedright a) Target Mach number $M$, (b) Target parallel ion flux $\Gamma_{t,\|}$, (c) Target density $n_t$, and (d) Global accuracy on particle, momentum and energy balances $\epsilon_{tot}$, plotted as a function of the upstream density $n_u$, for different forms of the Bohm condition and different values of the numerical viscosity $\nu_{num}$.}
\end{figure}

The base case, section \ref{sec:basecase}, leads to naturally supersonic flows ($M\approx1.7$) at the target after roll-over, Figure \ref{fig:cs_vs_above_s}a. When enforcing $M=1$ at the target in such conditions, the code showed numerical difficulties to converge with satisfactory particle and energy balances, with $\epsilon_{tot}$ (equation (\ref{eq:epsilon_tot})) reaching up to $\approx 12\%$, Figure \ref{fig:cs_vs_above_s}d. This is due to the presence of a very sharp velocity gradient required to slow down the flow to $M=1$. This is alleviated by the introduction of some numerical viscosity $\nu_{num}$ (equation (\ref{eq:par_visc})). Figures \ref{fig:cs_vs_above_s}b and \ref{fig:cs_vs_above_s}c plot the evolution of various target quantities for three different values of $\nu_{num}$. The effect of this artificial viscosity on $T_{e,t}$ (not shown), $n_t$ and $\Gamma_{t,\|}$ is modest, but it greatly improves the code convergence, with $\epsilon_{tot}$ reduced to a maximum of $\approx 1.7\%$ and $\approx 0.23\%$ for $\nu_{num} = 1.0\times10^{-5} [a.u.]$ and $\nu_{num} = 5.0\times10^{-5} [a.u.]$, respectively, Figure \ref{fig:cs_vs_above_s}d. We note here that it is not yet clear whether the occurrence of such sharp velocity gradient is a consequence of the choice of parameters for the base case, or a general observation. 

For both target boundary conditions $M>1$ and $M=1$, $\Gamma_{t,\|}$ is very similar, Figure \ref{fig:cs_vs_above_s}b. Similarly, the target temperature $T_{e,t}$ is unaffected (not shown). However, as expected, a difference arises in the target density $n_t$ and parallel velocity $u_{\|,t}$. The case with $M=1$ features a lower $u_{\|,t}$ and a higher $n_t$ compared to the $M>1$. Since $\Gamma_{t,\|}$ remains similar across these different cases, we conclude that enforcing the strict equality $u_\| = c_s$ leads to a redistribution of  $\Gamma_{t,\|}$ between its velocity and density contributions. This could have some implications in simulation codes that enforce $u_\| = c_s$, by promoting higher density at the target. Since many atomic source and sink terms scale with $n_e$, or even $n_e^2$, this will ultimately influence the particle, momentum, and energy balance of the system. 

%As a side remark, we note that in the $M>1$ case, the transition to super-sonic flows at the target is actually noticeable on $n_t$, with the presence of a ``bump" near $n_u \approx 1.9\times10^{19}~\mathrm{m^{-3}}$.

\section{Advanced studies enabled by SPLEND1D}\label{sec:change_model}
SPLEND1D opens up a large number of possible SOL and detachment studies, such as the investigation of the effect of the total flux expansion, of the parallel connection length, of in-out power sharing, or dynamical behavior. In the following sections, we present some example studies on the role of different neutral models, ion vs electron heating ratios, and heat pulses effects on the SOL plasma. 

\subsection{Neutral model}
The choice of the neutral model, and the neutral temperature implementation, can affect the plasma dynamics, primarily through the momentum and energy volumetric source terms. The diffusive model (equations (\ref{eq:nn_diff}), (\ref{eq:dn})) can be implemented with either $T_n=T_i$ or with a constant, imposed value of $T_n$ (in this section we set $T_n=1~$eV to model cold neutrals), while the advective model (equations (\ref{eq:neutral_dens_}), (\ref{eq:neutral_vp}), (\ref{eq:neutral_vt})) can be implemented with either $T_n=T_i$ or with the evolution of $T_n$ following equation (\ref{eq:neutral_energy}). 

\begin{figure*}
    \centering
    \includegraphics[width=\linewidth]{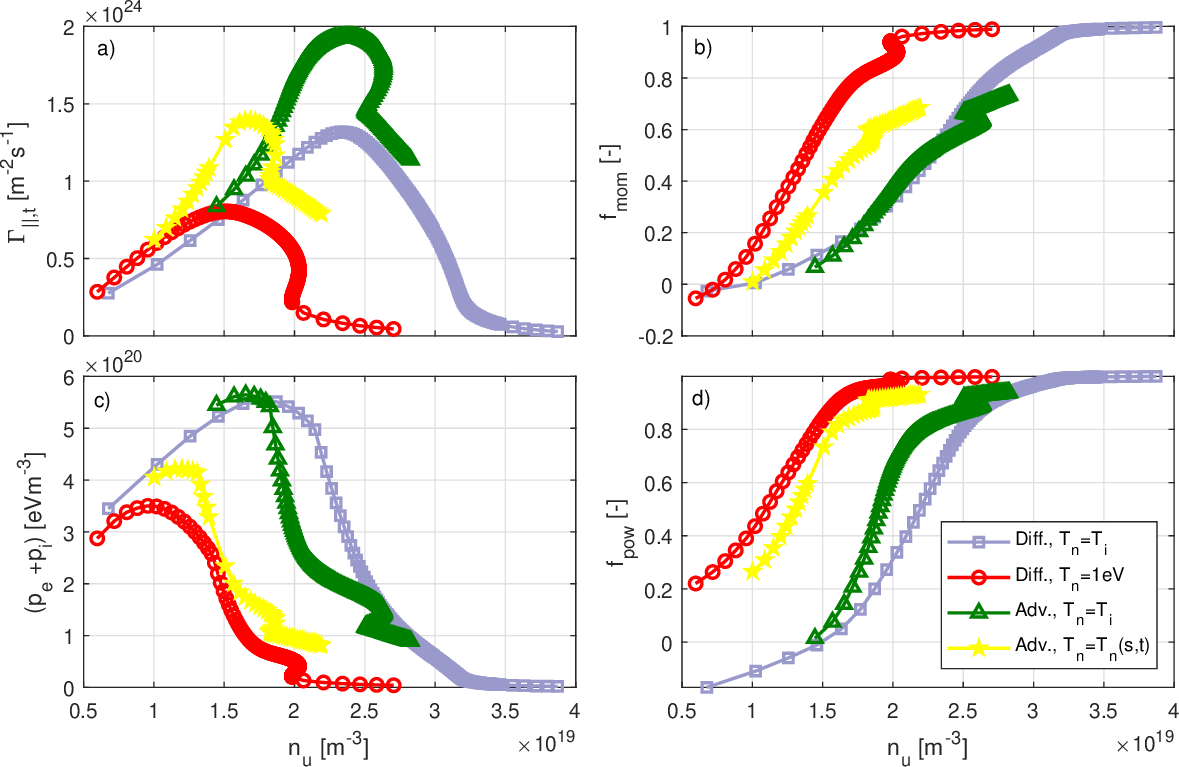}
    \caption{\raggedright (a) Target ion current, (b) momentum loss factor, (c) total plasma pressure, and (d) power loss factor, as a function of $n_u$ for a range of different neutral models/options implemented in SPLEND1D. The base case density ramp is performed using the following neutral models: diffusive model, with $T_n=T_i$; diffusive model, with $T_n=1~$eV; advective model, with $T_n=T_i$; advective model, evolving $T_n$ according to equation (\ref{eq:neutral_energy}).}
    \label{fig:neutralmodels}
\end{figure*}

Figure \ref{fig:neutralmodels} shows the main differences in the results of the base case simulations for each neutral model outlined above. In the advective model with a self-consistent $T_n$, and for the diffusive model with cold neutrals ($T_n=1~$eV), we see an earlier target ion current rollover, along with a lower target plasma pressure, compared to implementing $T_n=T_i$. The cold neutrals facilitate detachment by increasing both the power loss factor with respect to the cases with $T_n=T_i$, as a result of increasing the power transferred through charge exchange reactions, by increasing the energy transferred from the ions to neutrals. This, in turn, lead to an increase of the momentum losses by promoting enhanced charge-exchange reaction and reduced ionization. 

The advective model, in comparison to the diffusive model, features an additional neutral transport mechanism, the advective cross-"flux-tube" transport. For a fixed $\tau_N$, as shown in Figure \ref{fig:neutralmodels}, the advective model displays a much weaker rollover. The total momentum and power loss factors shown in Figure \ref{fig:neutralmodels} also differ between the models. 

\subsection{Independent ion and electron temperatures}
SPLEND1D can either solve a single energy equation, equation (\ref{eq:sf_energy}), assuming a proportionality relation between the ion and electron temperatures, or two separate energy equations for the electrons and ions, equations (\ref{eq:eenergy}) and (\ref{eq:ienergy}). The base case presented in section \ref{sec:basecase} has so far employed the $T_e = T_i$ assumption. In this section, we relax this constraint by enabling both temperatures to be independent. It is well known that, at low collisionality, ion and electron temperatures can be different \cite{stangeby2000plasma}. 

\begin{figure*}
    \centering
    \includegraphics[width=0.75\linewidth]{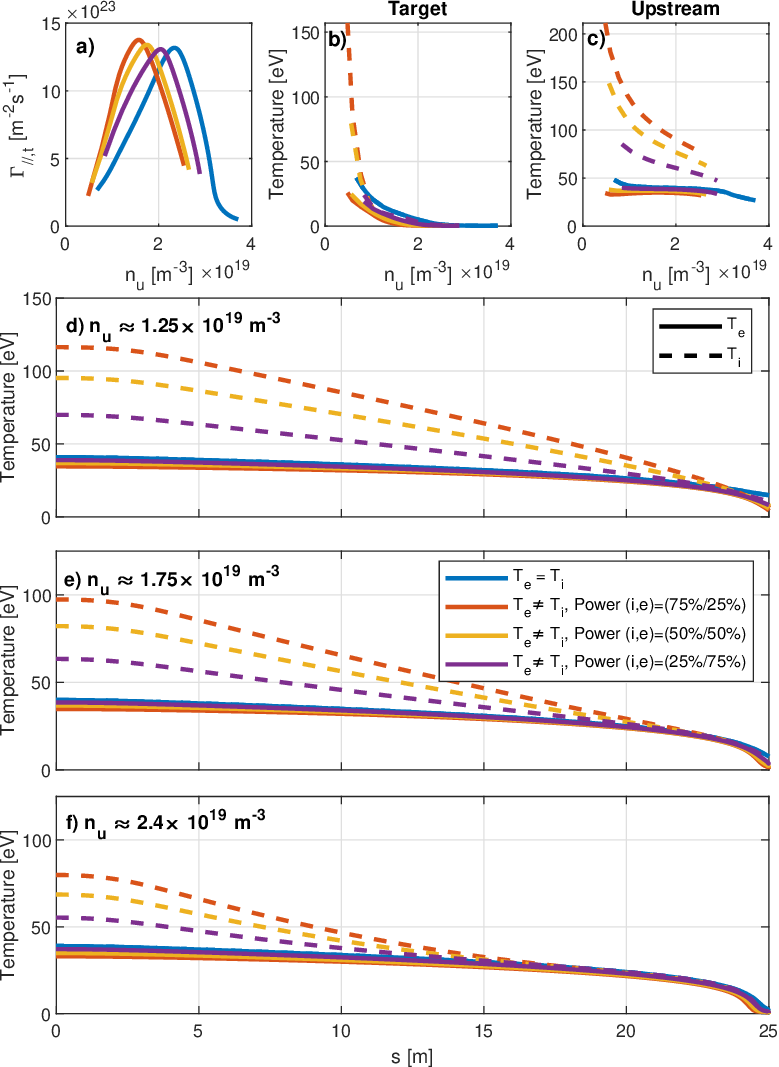}
    \caption{\raggedright (a) $\Gamma_{\|,t}$ as a function of upstream density $n_u$, for the base case with $T_i = T_e$ (blue), the base case with $T_i \ne T_e$ and a 75/25 power sharing between ions and electrons (red), the base case with $T_i \ne T_e$ and a 50/50 power sharing between ions and electrons (yellow) and the base case with $T_i \ne T_e$ and a 25/75 power sharing between ions and electrons (purple) (b) Target ion (dashed) and electron (solid) temperatures for the same cases as in Panel (a). (c) Upstream ion (dashed) and electron (solid) temperatures for the same cases as in Panel (a). Panels (d)-(e)-(f) Ion (dashed) and electron (solid) temperature profiles along the flux tube for $n_u = 1.25 \times 10^{19}~\mathrm{m^{-3}}$ (d), $n_u = 1.75 \times 10^{19}~\mathrm{m^{-3}}$ (e), and $n_u = 2.4 \times 10^{19}~\mathrm{m^{-3}}$ (f).}
    \label{fig:ti_vs_te}
\end{figure*}

To set up the simulations, we split the power source between the $H_e$ and $H_i$ terms (equations (\ref{eq:eenergy})-(\ref{eq:ienergy})), with either $H_e=H_i$ (50/50 split of the total input power between electrons and ions), $H_e=3H_i$ (75/25 split of the input power between electrons and ions), or $3H_e=H_i$ (25/75 split of the input power between electrons and ions). In all cases presented in this section, the total power injected in the system is kept constant, as well as all the parameters presented in table \ref{tab:basecase_parameters}. Similarly to section \ref{sec:detachment_basecase}, we perform density ramps by scanning $\bar{H_n}$. Allowing $T_i \neq T_e$ affects the roll-over threshold, Figure \ref{fig:ti_vs_te}a, and leads to typically higher $T_i$ than $T_e$, Figures \ref{fig:ti_vs_te}b) and \ref{fig:ti_vs_te}c). This is easily explained by the lower heat conduction coefficient of the ions, compared to the electrons, equations (\ref{eq:spitzer_heatcond_electron})-(\ref{eq:spitzer_heatcond_ion}). While the convective heat-flux and the equipartition term decrease this difference in transported heat flux. If the fraction of input power carried by the electrons is increased, the differences between $T_i$ and $T_e$ decreases, as expected, but remain significant. As density is increased, so does the collisionality and hence the equipartition term (equation (\ref{eq:equipartition})). The difference between $T_i$ and $T_e$ becomes negligible near the target (where density is high and temperature low), Figure \ref{fig:ti_vs_te}b, and reduces at the upstream location, Figure \ref{fig:ti_vs_te}c. This is also evident when looking at the temperature profile along the flux tubes, Figure \ref{fig:ti_vs_te}d, Figure \ref{fig:ti_vs_te}e and Figure \ref{fig:ti_vs_te}f which show profiles of $T_i$ and $T_e$ at increasing densities. Clearly, while in the attached case (Figure \ref{fig:ti_vs_te}d), the two temperatures are strongly different, they get closer to each other as density increases, Figure \ref{fig:ti_vs_te}e and Figure \ref{fig:ti_vs_te}f.

\subsection{Time-dependent simulations}\label{sec:timedep}
\begin{figure}
\centering
\includegraphics[width=1.00\linewidth]{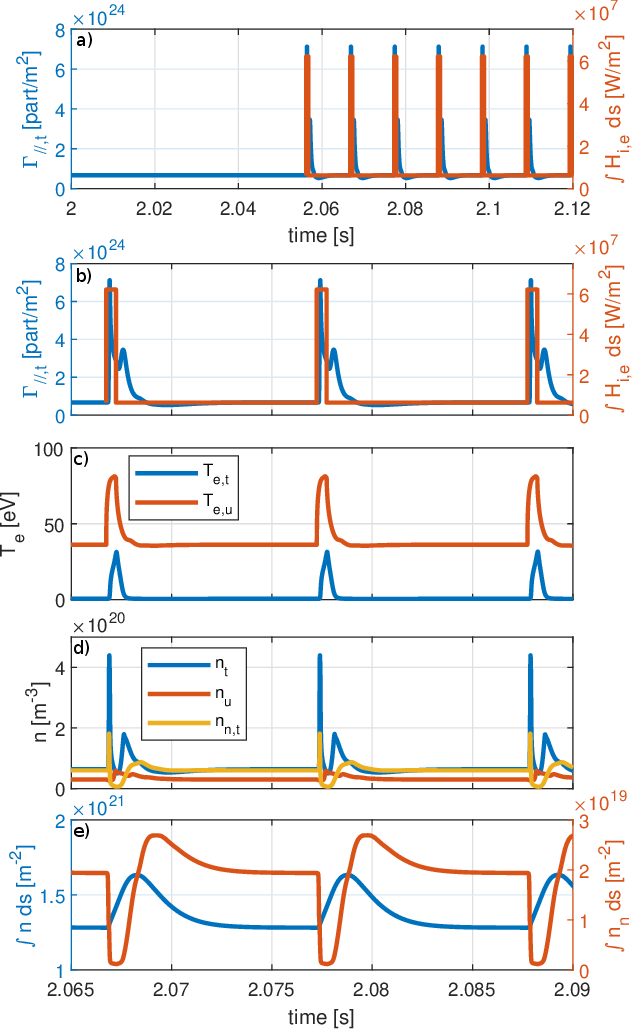}
\caption{\label{fig:time_depend_heat_pulses}\raggedright (a) Evolution of the target particle flux $\Gamma_{\|,t}$ (left axis, blue) and imposed heat-source integral $\int {H}_{i,e}~\mathrm{d}s$ as a function of time, for a sequence of 10 heat pulses, with (b) a zoomed version of panel (a). (c) time-evolution of the target electron temperature ($T_{e,t}$) and upstream electron temperature ($T_{e,u}$), (d) Time-evolution of the target density ($n_t$), upstream density ($n_u$) and target neutral density ($n_{n,t}$) and (e) time-evolution of the integrated plasma ($\int n~\mathrm{d}s$) and neutral ($\int n_n~\mathrm{d}s$) densities along the flux tube.}
\end{figure}

As mentioned in section \ref{sec:numerics}, SPLEND1D solves the 1D Braginskii equations as a time-dependent problem. Hence, in addition to finding steady-state solutions, as discussed in earlier sections, it is also possible to use SPLEND1D to explore the dynamics of the system. In this section, we briefly highlight such a possible study enabled by SPLEND1D. We perform a time-dependant simulation, based on a converged, detached simulation of the base-case, with $n_u = 3.0\times 10^{19}~\textrm{m}^{-3}$. We then introduce a sequence of short ($500~\mathrm{\mu s}$) pulses during which the heat-sources, $H_e$ and $H_i$, are amplified by a factor 10, before being relaxed to their initial values for $10~\mathrm{ms}$, Figure \ref{fig:time_depend_heat_pulses}a. During each pulse, we observe a strong increase of the target parallel particle flux, $\Gamma_{\|,t}$, Figure \ref{fig:time_depend_heat_pulses}b, together with a strong increase of the target electron temperature $T_{e,t}$ Figure \ref{fig:time_depend_heat_pulses}c, a sign that the plasma is reattaching during these events. 

The target plasma and neutral densities are also strongly affected by the heat pulses. In particular, after a short increase of the target neutral density $n_{n,t}$ (likely due to an increased recycling caused by the increased $\Gamma_{\|,t}$), $n_{n,t}$ drops below its steady-state value, due to the ionization of most of the neutrals present in the system, as evidenced in Figure \ref{fig:time_depend_heat_pulses}e) by the strong decrease of the integrated neutral density in the flux tube, associated with an increase of the integrated plasma density. This leads to an increase of the plasma upstream density $n_u$ and a complex dynamics of the plasma target density $n_t$, which, after an initial rise and drop, peaks again before relaxing to its steady-state value. These results highlight how SPLEND1D can be used for time-dependent simulations. 

\section{\label{sec:conclusion} Conclusion}
This paper presented the SPLEND1D code: a 1D plasma fluid model that solves the Braginskii equations projected along a flux tube, together with a simple neutral fluid model. We have highlighted the key features of SPLEND1D, including its flexibility in terms of included SOL physics (magnetic field norm and pitch angle variation along the flux tube, dependent or independent electron and ion temperatures, possibility to enable, disable, or rescale any physical term contributing to the equations, etc.), numerical accuracy, and high computational speed. We believe this makes SPLEND1D a valuable tool for studying the complex dynamics of a divertor plasma interacting with a neutral gas and in contact with a wall. The SPLEND1D code was then used to investigate the physics of detachment in a reference simulation scenario. The code outputs the individual contributions of each term constituting its particle, momentum, and energy equations, allowing for the results to be easily interpreted. The roll-over of the target ion flux, and the onset of detachment, was found to be mainly due to momentum losses owing to charge-exchange reactions, and power losses owing to impurity radiation. As with all reduced models, SPLEND1D comes with significant assumptions and free parameters. In this paper, we have shown how they can influence simulation results. SPLEND1D is currently being used to interpret TCV experiments, in particular those related to alternative divertor configurations, where it is being employed to elucidate the role of parallel connection length on the onset of detachment, as well as the impact of total flux expansion on detachment threshold and SOL parallel profiles. 

In the future, we plan on continuing the development of the SPLEND1D model. From a numerical perspective, it would be interesting to further improve the capabilities of SPLEND1D, for instance by adaptively refining the mesh depending on local gradients of the solution, which may prove useful in situations where the detachment ``front" is moving away from the target. From a physics perspective, it would be interesting to add the capability to handle multiple plasma and neutral species, such as molecules, which are known to play a significant role in the divertor dynamics \cite{Zhou_PPCF2022}. The addition of a collisional radiative model for impurities, to account for non-coronal effects in the simulations could be of interest. This could either be done with a trace assumption, or using a more advanced fluid closure such as the Zdhanov closure. 

\clearpage
\appendix
\section{Pressure-diffusion equation}\label{app:pressure_diffusion}
In this appendix, we show how equations (\ref{eq:neutral_vt}) and (\ref{eq:neutral_vp}) can be used to establish a pressure-diffusion equation for $\mathbf{V_n}$, which can then be incorporated in equation (\ref{eq:neutral_dens_}). This derivation is similar to the one presented in Ref. \cite{Horsten_NME2017}, and differs only by the presence of $\frac{1}{\tau_N}$ terms. Starting from equation (\ref{eq:neutral_vt}), we assume the neutral population to be at steady-state, such that the time-derivative can be removed. Further, we assume the neutral flow to be strongly subsonic, such that the inertia term $n_n m_n {V_n^\theta}^2$ can be neglected when compared to the the neutral pressure $p_n$. We simplify the expression by assuming that the variation of the magnetic field norm along the flux tube can be neglected. Under these assumptions, equation (\ref{eq:neutral_vt}) can then be simplified as
\begin{widetext}
\begin{equation}
     0 = -\frac{1}{\sin\alpha}\frac{\partial p_n}{\partial s} 	 {-m_inn_n\left< \sigma v\right>_{ion}{V_n^\theta}}   {+ m_in^2\left< \sigma v\right>_{rec}u_{\|}\sin\alpha} {- m_in_n n \left< \sigma v\right>_{CX}\left( {V_n^\theta} - u_{\|}\sin\alpha  \right)} - \frac{m_nn_nV_n^\theta}{\tau_n} \\
\end{equation}
\end{widetext}
leading to 
\begin{equation}
      n_n V_n^\theta = \frac{\left(n_n \left< \sigma v\right>_{CX}  + n\left< \sigma v\right>_{rec}\right)nu_{\|}\sin\alpha - \frac{1}{m_n\sin\alpha}\frac{\partial p_n}{\partial s}}{n\left< \sigma v\right>_{ion} +  n \left< \sigma v\right>_{CX} + \frac{1}{\tau_n}}
\end{equation}
which can readily be incorporated in equation (\ref{eq:neutral_dens_}). Similarly, and under the assumption that $n_n m_n {V_n^\theta}{V_n^\phi}$ is small, equation (\ref{eq:neutral_vp}) can be rewritten as
\begin{equation} 
{n_nV_n^\varphi} = \frac{\left(n\left< \sigma v\right>_{rec} + n_n \left< \sigma v\right>_{CX} \right) n u_{\|}\cos\alpha}{n\left< \sigma v\right>_{ion}+ n \left< \sigma v\right>_{CX} + \frac{1}{\tau_n}}
\end{equation}
The implementation and test of this formulation in SPLEND1D is, however, left for future work. 

\section{Contributions of momentum and power source terms to loss factors}\label{app:2PM_fmom_fpower}
The momentum and power source/sink terms can be calculated from the SPLEND1D model to evaluate the significance of each term, as shown in figure \ref{fig:IonizationVsRecombination}. Integrating the steady state form of the momentum equation (\ref{eq:sf_momentum}) along the flux tube length from the target to any upstream location, we can write the SPLEND1D equations in the form $p_{tot,u}-p_{tot,t}=f_{mom}{p_{tot,u}}$. We can then separate each of the contributing terms to find $f_{mom}=f_{mom}^{atomic}+f_{mom}^{visc}+f_{mom}^{\nabla_\|}$, where,
\begin{align}
f_{mom}^{atomic} &= \frac{1}{p_{tot,u}} \int^u_t S^{u}_{p,||} \mathrm{d}s \\
f_{mom}^{visc} &= \frac{1}{p_{tot,u}} \int^u_t \frac{4}{3}\left[ B^{3/2}\frac{\partial}{\partial s} \left( \eta_{||} B^{-2}\frac{\partial B^{1/2} u_{||}}{\partial s} \right) \right] \mathrm{d}s  \\
f_{mom}^{\nabla_\|} &= \frac{1}{p_{tot,u}} \int^u_t \frac{\partial B}{\partial s} \frac{m_i n u^2_{||}}{B} \mathrm{d}s.
\end{align}
$f_{mom}^{atomic}$ can be further decomposed as the sum of each atomic process included (ionization, recombination and charge-exchange, Equation (\ref{eq:momentum_term_expression}). Similarly to the momentum equation, each contributing terms in $f_{pow}$ can be separated, leading to $f_{pow}=f^{atomic}_{power}+f^{imp}_{power}$, where,
\begin{align}
f^{atomic}_{pow} \frac{q_{in}}{B_u} &=\int^u_t \frac{(S_e^E + S_i^E)}{B}\mathrm{d}s,  \\
f^{imp}_{pow}\frac{q_{in}}{B_u} &=\int^u_t \frac{S_{imp}^E}{B} \mathrm{d}s.
\end{align}
$f_{pow}^{atomic}$ can also be expanded as a sum of each atomic process included (equations (\ref{eq:atomics_ions_energy}) and (\ref{eq:atomics_electrons_energy})): ionization, recombination, charge-exchange and excitation. 

\begin{acknowledgments}
This work was supported in part by the Swiss National Science Foundation. This work has been carried out within the framework of the EUROfusion Consortium, via the Euratom Research and Training Programme (Grant Agreement No 101052200 — EUROfusion) and funded by the Swiss State Secretariat for Education, Research and Innovation (SERI). Views and opinions expressed are however those of the author(s) only and do not necessarily reflect those of the European Union, the European Commission, or SERI. Neither the European Union nor the European Commission nor SERI can be held responsible for them. 
\end{acknowledgments}

\section*{References}
\bibliography{1D_model}% Produces the bibliography via BibTeX.

\end{document}